\documentclass[apj,twocolappendix]{emulateapj}
\usepackage{apjfonts}

\begin{document}
\title{Ohmic Dissipation in the Interiors of Hot Jupiters}
\author{Xu Huang\altaffilmark{1} and Andrew Cumming\altaffilmark{2}}
\altaffiltext{1}{Department of Astrophysical Sciences, 4 Ivy Lane, Peyton Hall, 
Princeton University, Princeton, NJ 08544, USA;
email:xuhuang@princeton.edu}
\altaffiltext{2}{Department of Physics, McGill University, 3600 rue University, Montreal QC, H3A\ 2T8, Canada; email:cumming@physics.mcgill.ca}


\begin{abstract}
We present models of ohmic heating in the interiors of hot jupiters in which 
we decouple the interior and the wind zone by replacing the wind 
zone with a boundary temperature $T_{\rm iso}$ and magnetic field
$B_{\phi 0}$. 
Ohmic heating influences the contraction of gas giants in two ways: by direct 
heating within the convection zone, and by heating outside the convection zone 
which increases the effective insulation of the interior. We calculate these 
effects, and show that internal ohmic heating is only able to slow the 
contraction rate of a cooling gas giant once the planet reaches a critical 
value of internal entropy. We determine the age of the gas giant when ohmic 
heating becomes important as a function of mass, $T_{\rm iso}$ and induced 
$B_{\phi 0}$. With this survey of parameter space complete, we then adopt the 
wind zone scalings of Menou (2012) and calculate the expected evolution of gas 
giants with different levels of irradiation. We find that,with this
prescription of magnetic drag, it is difficult to inflate massive planets or 
those with strong irradiation using ohmic heating, meaning that we are unable 
to account for many of the observed hot jupiter radii. 
This is in contrast to previous evolutionary models that assumed that a 
constant fraction of the irradiation is transformed into ohmic power. 
\end{abstract}


\section{Introduction}

The large radii of many hot jupiters has been a puzzle ever since the
discovery of the first transiting planet HD~209458b (\citet{Charbonneau00}; 
see \citet{Baraffe2010} for a review). \citet{Showman2002a} pointed out that if 
a certain amount of the irradiation from the star ($\sim1\%$ of the incident 
stellar flux) can be deposited deep in the envelope of the planet (pressures 
of $\gtrsim 10$ bars), then the inflated radius can be explained. But the 
physical mechanism by which the required energy is transported into the 
interior of planet is still an open question. Several explanations have been 
proposed, such as a downward kinetic flux due to atmosphere circulation, or 
turbulent transport and shock heating in the flow 
\citep{Showman2002a,Youdin2010,Perna2012}, or tidal dissipation 
\citep{Bodenheimer2001,Bodenheimer2003}. But each of these has problems in 
accounting for all of the observed radii (e.g.~\citealt{Laughlin2011,Demory2011}).

\citet{B&S2010} suggested that ohmic heating could serve as the heat source in 
the interior of inflated hot jupiters. The idea is that the shearing of the 
planetary magnetic field by the wind driven in the outer layers of the planet 
by irradiation generates an induced current that flows inwards, dissipating 
energy by ohmic dissipation in deeper layers (see also \citealt{Liu2008}). 
\citet{Perna2010a,Perna2010b} also pointed out the possible importance of 
magnetic drag on the dynamics of the flow in the envelope, and found that a 
significant amount of energy could be dissipated by ohmic heating at depths 
that could influence the radius evolution of the planet. 

\cite{B&S2011} implemented ohmic heating in evolutionary models of gas giants 
and showed that the amount of inflation depends significantly on the amount of 
irradiation received by the planet (and therefore its equilibrium temperature 
$T_{\rm eq}$). They found that the radii of low mass planets could run away, 
increasing dramatically in response to ohmic heating and leading to evaporation 
of the planet. This same behavior was not observed in the recent study of 
\cite{Wu2012}, who find that ohmic heating could increase the radius of a 
planet that had already cooled, but only modestly. On the other hand, if ohmic 
heating operates early in the lifetime of a hot jupiter, \cite{Wu2012} find 
that contraction can be halted and large radii obtained, large enough to 
explain the observed radii of all except a few planets (see their Fig.~4).

Both the time evolution models of \cite{B&S2011} and \cite{Wu2012} calculate 
the profile of ohmic heating as $J^2/\sigma$ per unit volume inside the planet 
(where $J$ is the current density and $\sigma$ the electrical conductivity), 
but adjust the overall level of heating so that the efficiency $\epsilon$ --- 
the fraction of the irradiation that goes into ohmic heating --- is fixed at a 
level of $\epsilon\sim1$\%. In reality, the magnitude of the current that 
penetrates into the interior depends on how the flow in the wind zone 
interacts with the planet's magnetic field and the feedback from the magnetic 
field on the flow dynamics. \cite{Menou2012} considers scaling arguments for 
the atmospheric flows in a magnetized atmosphere, and argues that the 
efficiency $\epsilon$ must decline at large $T_{\rm eq}$ as magnetic drag 
limits the flow velocity in the atmosphere 
(see also \citealt{Perna2010b,Rauscher2012}). 

In this paper, we take a more general approach to the question of inflation 
due to ohmic heating, with the aim of understanding the different evolutions 
found by \cite{B&S2011} and \cite{Wu2012}, and incorporating the effect of 
magnetic drag, and therefore variable efficiency, on the evolution. We take a 
different approach by separately considering the planet interior and the wind 
zone. We first calculate the interior heating by replacing the wind zone with a 
boundary condition which specifies the toroidal field, or equivalently the 
radial current, at the base of the wind zone and the temperature there. This 
allows us to survey the parameters that influence the amount of ohmic heating 
and its effect on the evolution of the planet. In this way we go beyond the 
previous assumptions of constant heating efficiency. We then implement the 
scaling laws derived by \cite{Menou2012}, and show that indeed the efficacy of 
ohmic heating is reduced at high $T_{\rm eq}$ because of increased drag in the 
wind zone. In this way our time dependent models differ crucially from 
\cite{B&S2011} and \cite{Wu2012} in that we find that it is difficult to 
explain the observed radii of many hot jupiters with ohmic heating under the influence
of magnetic drag,  particularly those with large masses $M\gtrsim M_J$.
    
The plan of the paper is as follows. In \S \ref{Con}, we review the general 
mechanism of ohmic heating and how the internal current is calculated, giving 
some order of magnitude estimates for the total ohmic power. 
Next in \S \ref{Ty}, we present quasi-steady state models of gas giants 
undergoing ohmic heating as a function of their internal entropy $S$ and the 
induced magnetic field $B_{\phi 0}$ and temperature $T_{\rm iso}$ at the base 
of the wind zone. We then use these models to follow the time evolution of a 
given planet under the action of ohmic heating in \S \ref{evolution}. With this 
general survey of parameter space in hand, we then use the scalings derived by 
\cite{Menou2012} for the wind zone to calculate the evolution of observed hot 
jupiters and compare with observed radii (\S \ref{observation}). We discuss the 
limitations of our models and compare to other work in \S \ref{discuss}.


\section{The General Mechanism of Ohmic heating}
\label{Con}

In this section, we review the basic physics of ohmic heating, focussing on the 
generation of the induced field in the wind zone and radial current that 
penetrates into the planet interior (\S\ref{sec:current}). We then estimate the 
likely magnetic field strengths that can be generated in the wind zone and the 
resulting ohmic power available for inflation (\S\ref{AR}).

\subsection{Calculation of the induced field and current distribution }
\label{sec:current}

First we review the general picture put forward by \citet{B&S2010} (See
also \citep{Perna2010a}). 
Due to strong irradiation from the host star, the hot jupiter has a 
thermally-ionized atmosphere, coupling the magnetic field and the atmospheric 
flow. The magnetic field could be either the intrinsic planetary magnetic 
field generated by a dynamo in the deep interior, or the external field from 
the host star. In either case, the evolution of the magnetic field in the wind 
zone is governed by the induction equation 
\begin{equation}
\frac{\partial{\vec{B}}}{\partial{t}}=-\nabla\times\eta(\nabla\times{\vec{B}})+\nabla\times(\vec{v}\times{\vec{B}})
\label{eq:induc}
\end{equation}
where $\vec{v}$ is the wind velocity and $\eta$ is the magnetic diffusivity of 
the atmosphere. Assuming a steady state, and that the planet magnetic field in 
the outer layers is well-represented by a curl-free dipole field 
$\vec{B}_{\rm dip}$, the induced magnetic field $\vec{b}$ is given by
\begin{equation}
\nabla\times\eta(\nabla\times{\vec{b}})=\nabla\times(\vec{v}\times{\vec{B}_{\rm dip}}).
\label{eq:beq}
\end{equation}
If the wind is predominately a zonal flow $\vec{v}=v_\phi \hat{\phi}$, the 
induced field is toroidal, and will penetrate into the interior of the planet, 
with associated poloidal currents that close in the interior given by 
$\vec{J}=(c/4\pi)\nabla\times{\vec{b}}$. The internal toroidal field is given 
by
\begin{equation}\label{eq:beqinterior}
\nabla\times(\eta\nabla\times{\vec{b}})=0
\end{equation}
below the wind zone where velocities are negligible. For a given poloidal 
current $J$, the local ohmic dissipation rate is $P_{\rm ohm}=J^2/\sigma$, 
where $\sigma$ is the electrical conductivity, related to the magnetic 
diffusivity by $\eta=c^2/4\pi\sigma$.
\begin{figure}
\centering
\hbox{ 
\includegraphics[angle=0,width=\linewidth]{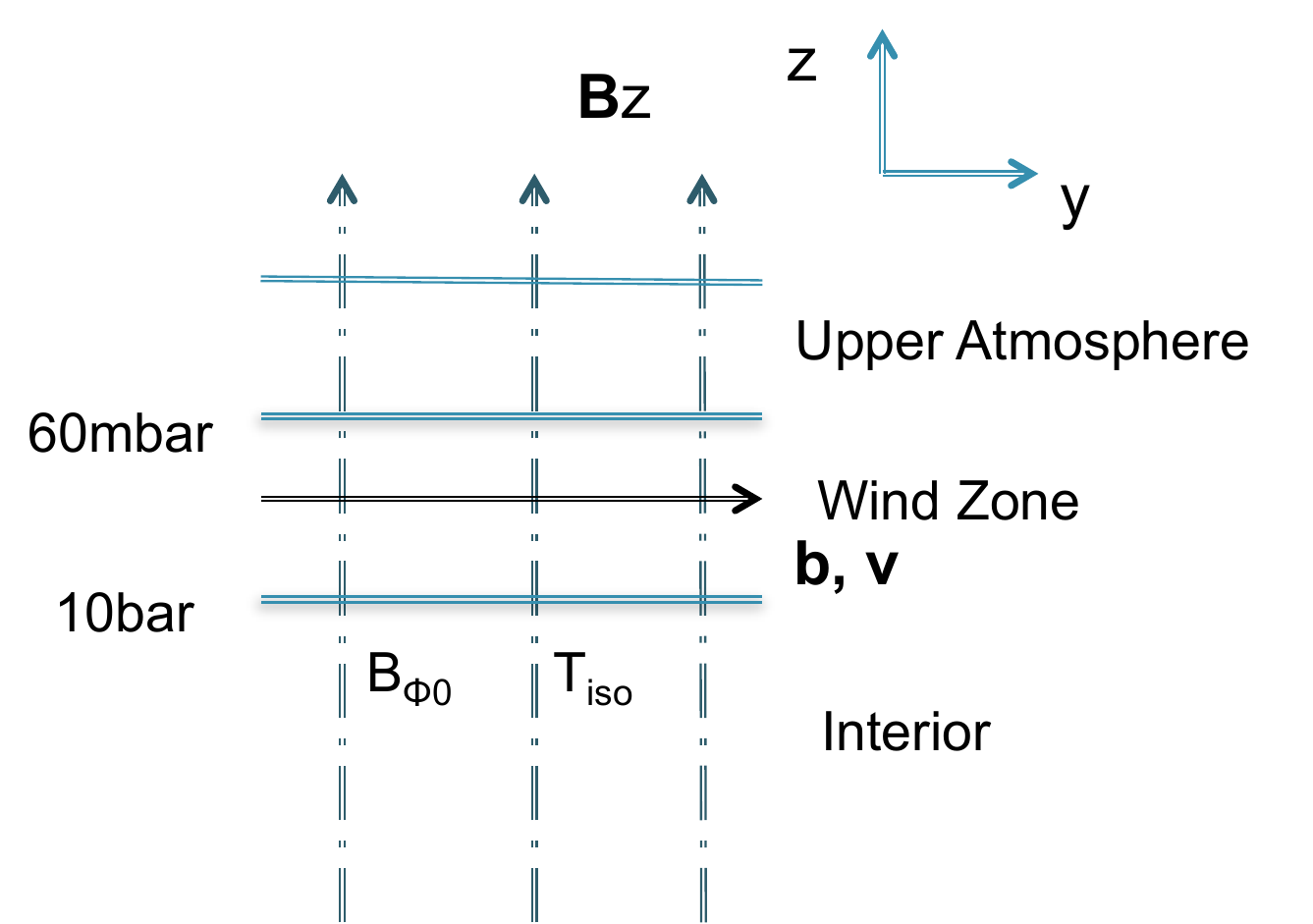}
}
\caption{
The general mechanism of ohmic heating illustrated in a plane-parallel model. 
A vertical field $B_z$ is sheared by the wind $v_y$ in the wind zone. 
An induced field $b$ is produced by the shear that penetrates into the interior.
We use the temperature $T_{\rm iso}$ and induced magnetic field
$B_{\phi 0}$ at the base of the wind zone as boundary conditions for our 
interior solutions.
\label{f:ppmodel}
}
\end{figure}

Some intuition about the solution can be obtained by considering a plane
parallel model, which we illustrate in Figure~\ref{f:ppmodel}. There we
divide the planet into three layers, representing the outermost
isothermal layer, with pressures lower than $60\ {\rm mbars}$, the
wind zone, between $60\ {\rm mbars}$ and $10\ {\rm bars}$ and the 
interior of the planet. The vertical field $B_z$ is sheared by the wind with 
velocity $v_y(z)\exp{(ikx)}$. Focusing on the innermost layer representing the 
planet interior, and assuming constant conductivity, equation 
(\ref{eq:beqinterior}) gives an induced field 
$\vec{b}\propto\exp(-{kz}+ikx)\hat{y}$ 
and associated vertical current $4\pi j_z/c=ikb$. 
Both the field and current decrease exponentially in the interior on a length 
scale $2\pi/k$. When the variation of conductivity with depth is included, we 
must solve 
\begin{equation}
\frac{d^2b}{dz^2}-k^2b+\frac{d\eta}{dz}\frac{db}{dz}=0,
\label{eq:ppeq}
\end{equation}
in which case the thickness of the penetration depth depends on the length 
scale on which the conductivity varies. 

This simple model illustrates that the interior solution depends on three 
factors: the geometry of the shearing velocity in the wind zone (which is 
described by $k$ in this simple model), the magnitude of the induced field at 
the base of the wind zone, and the profile of the electrical conductivity in 
the interior. The approach that we pursue in this paper is to consider the 
first two factors as boundary conditions on the interior. We choose to 
parametrize our models by specifying the toroidal magnetic field at a 
pressure of 10 bars, which we will refer to as $B_{\phi 0}$. 

To calculate the distribution of currents inside the planet in detail, we 
consider a simple geometry with a dipole field and a zonal flow 
$\vec{v}=v_0\sin{\theta}\,\hat{\phi}$. 
To solve equation (\ref{eq:beqinterior}) in spherical coordinates we 
write the induced toroidal field:
\begin{equation}
\vec{B_{\phi}}=\frac{g(r)}{r}\sin{\theta}\cos{\theta}\hat{\phi}, 
\end{equation}
in which case the radial dependence part of the field is given by 
\begin{equation}\label{eq:g}
g''(r)-\frac{d\ln\sigma}{dr}g'(r)-l(l+1)\frac{g(r)}{r^2}=0
\end{equation}
where $l=2$ is the index of associated Legendre polynomial $P_l^1$. 
Having found $g(r)$ and therefore $B_\phi(r)$, the currents are 
determined by Amp\`ere's law 
$\vec{J}=(c/4\pi)\nabla\times\vec{B}$. The ohmic power in the interior is
\begin{equation}\label{eq:Pint}
P=\int\frac{J^2}{\sigma}\,dV\sim4\pi\int\frac{\langle{J}\rangle^2}{\sigma}r^2\,dr,
\end{equation}
where $\langle{J}\rangle=\langle{J_r^2}+J_\theta^2\rangle^{1/2}$ is the 
effective angle averaged current at radius $r$.
To get some feeling for the dependence of the field and current on depth, we 
can consider a power-law dependence $\sigma\propto r^\alpha$. In that case, 
the solution is $B_\phi\propto r^\beta$ with 
\begin{equation}
\beta=\frac{(\alpha-1)+\sqrt{(1+\alpha)^2+24}}{2},
\label{eq:gr}
\end{equation}
and $\langle{J}\rangle\propto r^{\beta-1}$. For more complex wind geometries, 
which involve $l>2$, the solution is 
$\vec{B_\phi}\propto r^lP_l^1(cos\theta)$ 
for constant conductivity. For example, this indicates that more zonal 
jets in the wind zone implies a shallower penetration depth for the 
induced field.

In fact, as we argue in the next section (\S \ref{AR}), the internal heating is 
dominated by the lowest densities, since the local heating rate is inversely 
proportional to the electrical conductivity which increases rapidly with 
increasing pressure. This means that the current $J$ can be taken as a 
constant without making a significant error in the heating profile. We have 
confirmed this by comparing constant current solutions with detailed solutions 
of equation (\ref{eq:g}). 

For the constant current case, we compute the current as 
\begin{equation}\label{eq:JR}
J={c\over 4\pi}{B_{\phi 0}\over R_J},
\label{eq:Jr}
\end{equation}
where $R_J=7\times 10^9\ {\rm cm}$ is the radius of Jupiter, 
$B_{\phi 0}=B_{\phi} (r,p=10 {\rm bars})$. This means that 
there is a direct mapping between the chosen value of $B_{\phi 0}$, the radial 
current inside the planet $J_r$, and the local heating rate, taken to be 
$J_r^2/\sigma(r)$ per unit volume. Note that we do not take into account the 
averages over angle in equation (\ref{eq:Pint}) nor the true radius of the 
planet, and so our value of $B_{\phi 0}$ for a given amount of internal heating 
could be a factor of a few of the toroidal field in a model which 
self-consistently includes both the wind zone and the interior. Instead, our 
parameter $B_{\phi 0}$ should be interpreted as a measure of the internal 
heating (given by eqs.~[\ref{eq:JR}] and then $J_r^2/\sigma$ locally). 

\subsection{Magnitude of the induced field and ohmic power}
\label{AR}

It is useful to estimate the expected magnitude of ohmic heating and how it 
scales with parameters such as planet mass $M$. The first step is to use 
equation (\ref{eq:beq}) to estimate the expected strength of the induced field
by dimensional analysis, 
\begin{equation}
\frac{b}{B_{\rm dip}}=R_M=\frac{4\pi\sigma{H}v}{c^2},
\label{eq:RM}
\end{equation} 
or
\begin{equation}
b=B_{\rm dip}\left(\frac{\sigma}{10^7\,{\rm
s}^{-1}}\right)\left(\frac{H}{0.01\,R_J}\right)\left(\frac{v}{1\,{\rm
km}\,{\rm s}^{-1}}\right),
\label{eq:nbphi}
\end{equation}
where $v$ is an average wind speed and $\sigma$ a typical value of electrical 
conductivity in the layer\footnote{We give the electrical conductivity in cgs 
units here. Note that the conversion to SI units is 
$1\ \mathrm{S\ m^{-1}}=9\times 10^{9}\ {\rm s}^{-1}$.}, and we take the 
vertical length scale to be the pressure scale height $H$.

In this paper, we take $B_{\rm dip}=10 {\rm G}$ as a standard value. Typical 
dipole field strengths for hot jupiters have been estimated from scalings with 
planet parameters (see \citealt{Trammell11} for a detailed summary and 
discussion). \citet{Lavega2004} argued that the field is generated by the 
dynamo action in the metallic region as in Jupiter \citep{Stevenson83}, with 
the field strength closely related to the rotation of the planet, 
$B\sim(\rho\Omega\eta)^{\frac{1}{2}}$. This predicts that the field on typical 
hot jupiters should be a factor of few smaller than that on the Jupiter, with 
a typical value of equatorial magnetic field $B_{\rm eq}\sim5 {\rm G}$. 
However, \citet{Christensen09} argue that the field instead scales with the 
heat flux escaping from the conductive core at large enough rotation rate, 
giving $B\sim(\rho{F}_{\rm core}^2)^{\frac{1}{3}}$. This gives a field 
strength an order of magnitude larger than estimated with the previous method. 

Given an induced field $b$, we can then estimate the expected ohmic power per 
unit mass, $P_m=\langle{J}\rangle^2/(\rho\sigma)$. Approximating the angle 
averaged current at the base of the wind zone using the constant current case 
as equation (\ref{eq:Jr}) gives
\begin{equation}\label{eq:pohm}
P_m=10^{-1}\ \mathrm{erg\ g^{-1}\ s^{-1}} \left(\frac{B_{\phi 0}}{10 \mathrm{G}}\right)^2 \left(\frac{\sigma}{10^{6}\ \mathrm{s}^{-1}}\right)^{-1} \left(\frac{\rho}{10^{-4}\ \mathrm{g\ cm^{-3}}}\right)^{-1}.
\end{equation}
Since $\sigma$ increases dramatically in the core of the planet, heating at low density dominates. If the conductivity profile scales as $\exp(-r/H)$ for example, where we take the lengthscale as the pressure scale height $H$, the total power in the interior is
\begin{eqnarray}\label{eq:pohmtotal}
P_{\rm ohm, total}&\approx &4{\pi}P_{\rm top}{R}^2\rho{H}
\\&=&10^{23}\,\mathrm{erg\ s^{-1}}\,\left(\frac{B_{\phi 0}}{10\ \mathrm{G}}\right)^2\,\left(\frac{\sigma_t}{10^{6}\ \mathrm{s}^{-1}}\right)^{-1}\times\nonumber\\&&\left(\frac{H}{0.01\ R_J}\right)\left(\frac{R}{R_J}\right)^2\nonumber \\
&=& 3\times10^{22}\ \mathrm{erg\ s^{-1}}\,\left(\frac{B_{\phi 0}}{10\ \mathrm{G}}\right)^2\,\left(\frac{\sigma_t}{10^{6}\ \mathrm{s}^{-1}}\right)^{-1}\times\nonumber \\&&\left(\frac{T}{1500\ {\rm K}}\right)\,\left(\frac{R}{R_J}\right)^4\,\left(\frac{M}{M_J}\right)^{-1}
\end{eqnarray} 
where the subscript $t$ indicates a quantity evaluated in the outermost regions 
of the interior just below the wind zone. Note that $H=\mathcal{R}T/g$, where 
we adopt $\mathcal{R}=3.64\times10^{7}\ \mathrm{erg\ g^{-1}\ K^{-1}}$ for a 
hydrogen molecule dominated composition with helium fraction $Y=0.25$. As we 
mentioned previously, because the total ohmic power is dominated by the heating 
at low pressure, it is not very sensitive to the radial profile of the current. 
The scaling in equation (\ref{eq:pohmtotal}) implies that 
$P_{\rm ohm}\propto 1/M$ when the radius and conductivity of planet is not 
strongly depend on mass, a scaling that we find in our numerical solutions. 
More massive planets have less ohmic power for a given $B_{\phi 0}$ and 
temperature $T_{\rm iso}$ at the base of the wind zone.

\section{Ohmic Heating as a Function of Internal Entropy}
\label{Ty}

In this section, we calculate the structure, luminosity, and ohmic heating 
profile for gas giants as a function of their internal entropy $S$. We will 
use these models in \S \ref{evolution} to follow the time evolution of the 
planet by following the decreasing entropy as the planet cools. In this 
approach, described by \cite{Hubbard77} (see also  \citealt{Fortney03} and 
\citealt{Arras06}), it is assumed that the convective turnover time is much 
shorter than the evolution time of the planet, so that the convection zone 
maintains an adiabatic profile as it cools and lowers its entropy. We also 
assume that the radiative envelope has a thermal timescale much shorter than 
the evolution time, so that the envelope is in thermal steady-state, carrying 
the luminosity emerging from the convection zone. The luminosity of the planet 
is then given by the radiative luminosity at the radiative--convective boundary,
\begin{equation}\label{eq:lum}
L={16\pi GcM_r aT^4\over 3\kappa p_c}\nabla_{\rm ad}
\end{equation}
where $p_c$ is the pressure at the convective boundary, $T$ is the temperature 
at that location, and $M_r$ is the enclosed mass. The evolution of the 
internal entropy is then given by
\begin{equation}\label{eq:dSdt}
\int\,T {dS\over dt}\,dm = M\bar{T}{dS\over dt} = -L(S)
\end{equation}
where we have assumed $dS/dt$ is constant across the convection zone and 
define the mass-averaged temperature $\bar{T}=\int\,T/M\,dm$.

The effect of ohmic heating appears in two places in this approach. The first 
is that an ohmic heating term must be added to the right hand side of equation 
(\ref{eq:dSdt}), 
\begin{equation}\label{eq:dSdtohm}
M\bar{T}{dS\over dt}=-L_c+\int{J^2\over \sigma}\,dV,
\end{equation}
where the integral is over the convection zone.
For a planet with a given entropy $S$, the luminosity at the 
top of the convection zone is fixed by the structure and is given by equation 
(\ref{eq:lum}). However, because some of this luminosity is now provided by 
ohmic heating, the cooling rate of the convection zone ($dS/dt$) is smaller. 
The second influence of ohmic heating is that it can change the temperature 
profile in the radiative zone, in particular by pushing the 
radiative--convective boundary to higher pressure and lowering the 
luminosity (eq.~[\ref{eq:lum}]). 

In this section, we include the first effect by calculating gas giant models 
without feedback from ohmic heating in the atmosphere (\S 3.1), and then 
include the feedback from ohmic heating in the radiative zone to include the 
second effect (\S 3.2). In \S 3.3, we summarize the results.


\begin{deluxetable*}{lcrrccc}
\tablewidth{0pc}
\tablecaption{Model Summary}
\tablehead{
\colhead{Model}  & \colhead{$R(R_J)$} &
\colhead{$B_{\phi0}({\rm G})$} & \colhead{$p_{\rm conv}({\rm bars})$} & 
\colhead{$P_{\rm ohm}({\rm erg\ s^{-1}})(p>p_{\rm conv})$} & 
\colhead{$P_{\rm ohm}({\rm erg\ s^{-1}})(p>10\ {\rm bars})$} & 
\colhead{$L_{\rm conv}(\mathrm{erg\ s^{-1}})$} 
}
\startdata
Isothermal & 1.25 & 10 & 62.8&$8.0\times10^{23}$& $1.2\times10^{25}$ &$1.3\times10^{26}$\\
Isothermal  & 1.25 & 100 & 62.8&$8.0\times10^{25}$& $1.2\times10^{27}$ &$1.3\times10^{26}$\\
Radiative   & 1.25 & 10 & 131.7 &$6.8\times10^{22}$& $2.8\times10^{24}$ &$7.7\times10^{25}$\\
Radiative   & 1.25 & 100 & 131.7 &$6.8\times10^{24}$& $2.8\times10^{26}$ &$7.7\times10^{25}$\\
RadiativeFB\tablenotemark{b}  & 1.25 & 10 &132.0&$6.7\times10^{22}$&  $2.8\times10^{24}$ & $7.7\times10^{25}$\\
RadiativeFB\tablenotemark{b} & 1.25 & 100 &176.4&$3.1\times10^{24}$&  $1.9\times10^{26}$ & $5.6\times10^{25}$
\enddata
\tablenotetext{a}{
Model computed with $S=8$, $T_{\rm iso}=1500\ {\rm K}$, and $M=0.96\ M_J$. We 
refer to this set of input parameters our standard model in the text.
}
\tablenotetext{b}{
Model including the feedback of ohmic heating in the atmosphere.
}
\tablenotetext{c}{
Both the cooling luminosity and the ohmic heating in the convective zone reduce while 
the convective zone boundary move towards deeper pressure due to the feedback in up
atmosphere.
}
\label{t:models}
\end{deluxetable*}

\subsection{Planet models without feedback}

We now make models of gas giants with given mass $M$ and central entropy $S$, 
and use them to calculate the ohmic dissipation in the planet, but without 
including the effect of ohmic heating on the planet structure. This allows us 
to calculate the ohmic heating within the convection zone and, by including 
this ohmic power in equation (\ref{eq:dSdt}), the corresponding slowing of the 
cooling rate. 

We present the detail microphysics in our planet model in the Appendix
\ref{sec:microphysics}. Here we only note the differences between our opacity 
and conductivity profiles and those used in other works. 
Our opacity profile use a different extrapolation method in the 
intermediate pressure range between $10^3-10^5 {\rm bars}$ from 
\cite{Paxton11}, who take the opacity for $\log R>8$ (where $R=\rho/T_6^3$ 
is used in the opacity tables) to be a constant set by the value at $\log R=8$. 
As far as we are aware, opacity calculations for this intermediate pressure 
range have not been carried out. For planets undergoing a large amount of 
internal heating, the convection zone boundary moves into this region, and 
so knowing the opacity there is important for understanding the location of 
the convection zone boundary. As we describe later, this controls the 
contraction rate of the cooling planet.
Our potassium conductivity profile (equation [\ref{eq:sigv}]) is the same as 
used by \cite{Perna2010a}, but is different from \cite{B&S2010}, who 
use a electron-neutron cross-section of 
$\pi (7.2\times 10^{-9}\ {\rm cm})^2=1.6\times 10^{-16}\,{\rm cm^2}$ rather 
than $10^{-15}\,{\rm cm^2}$ \citep{Draine83} and a slightly different thermal 
averaging factor for the velocity. The resulting difference is that the 
conductivity of \cite{B&S2010} is a factor of 9 times larger than our 
conductivity. 

The planet model is calculated by integrating outwards from the center, 
following the convective adiabat until it intersects a radiative zone 
extending inwards from the surface. When calculating the cooling curve of an 
irradiated gas giant, a reasonable approximation is to take the outer 
radiative zone of the planet to be isothermal. However, because ohmic heating 
is very sensitive to pressure, a small error in the determination of the 
convective boundary results in a much larger error in the ohmic power. To 
illustrate this, we have calculated models with an isothermal radiative layer 
and with a radiative layer that is in thermal equilibrium and carries a 
constant luminosity equal to the luminosity from the convection zone.

\begin{figure}
\epsscale{1.1}
\plotone{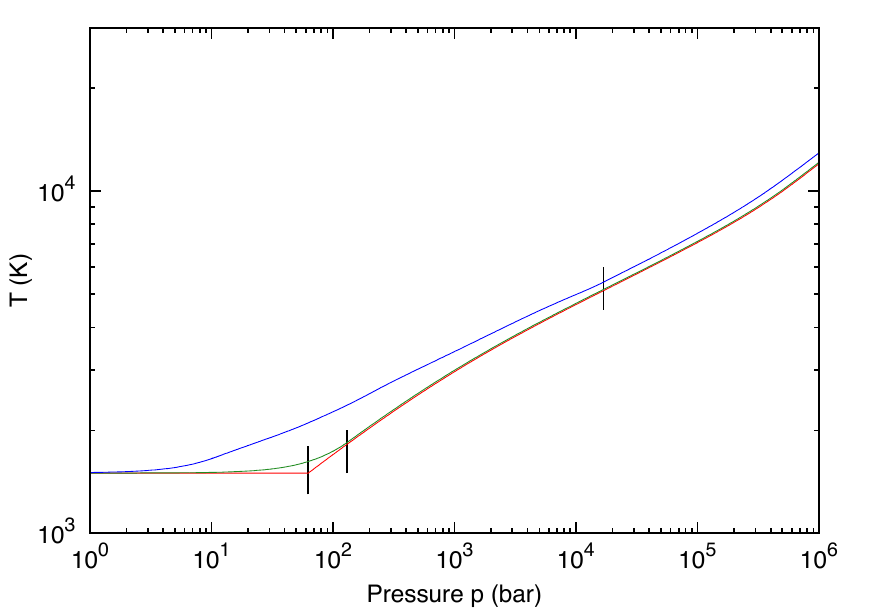}
\epsscale{1.15}
\plotone{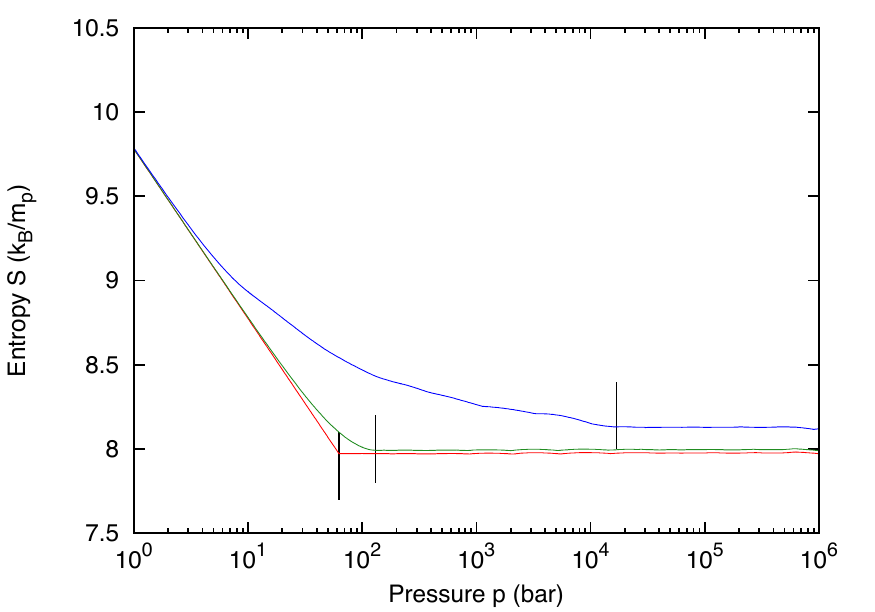}
\caption{
The temperature profile (top panel) and entropy profile (bottom panel) for 
different treatments of the outer layers, either isothermal (red curve), 
radiative (green curve), or radiative including ohmic heating (blue curve). 
For the ohmic heating model, we take $B_{\phi 0}=1000\ \mathrm{G}$. 
In each case, the radiative--convective boundary is marked with a vertical bar. 
Note that we do not show the entire structure, but focus on the lower 
pressures to illustrate the differences in the position of the 
radiative--convective boundary between models.
\label{f:structure}
}
\end{figure}

\begin{figure}
\epsscale{1.2}
\plotone{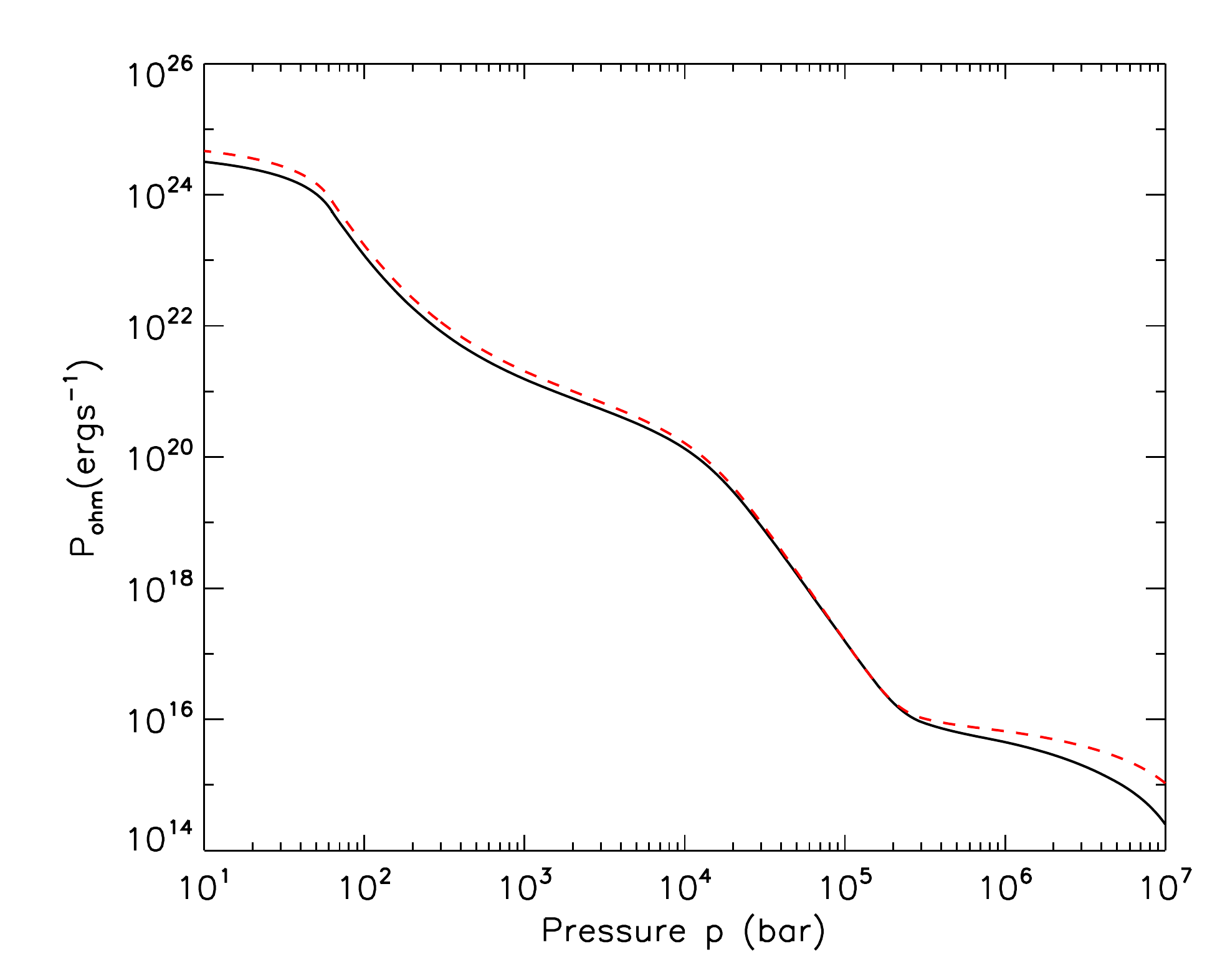}
\caption{
The cumulative ohmic power against pressure for the same planet parameters as 
in Figure~\ref{f:sigma}, with $B=10 {\rm G}$ (radiative model, no feedback). 
The solid curve uses a current profile calculated by solving 
equation~(\ref{eq:g}); the red dashed curve assumes a constant current with 
depth.
\label{f:solution}
}
\end{figure}


For the isothermal case, we integrate 
\begin{eqnarray}
\frac{dm}{dr}&=&4\pi\rho{r^2}\\
\frac{dp}{dr}&=&-\rho\frac{Gm}{r^2} \\
\frac{dT}{dr} &=&\frac{T}{P}\nabla\frac{dp}{dr}, 
\label{eq:struct}
\end{eqnarray}
outwards from the center, taking 
$\nabla=\nabla_{\rm ad}$ for $T>T_{\rm iso}$ (adiabatic interior) and 
$\nabla=0$ for $T<T_{\rm iso}$ (isothermal layer). 
In the non-isothermal case, we take 
\begin{eqnarray}
\nabla&=&\min{(\nabla_{\rm rad},\nabla_{\rm ad})} \\
\nabla_{\rm rad}&= &\frac{3{\kappa}L}{16{\pi}cGM}\frac{p}{aT^4}.
\end{eqnarray}
We calculate a model for a given $M$ and $S$ by integrating outwards
from the center and inwards from the surface to a matching pressure
$p=30\ {\rm kbars}$. For the outwards integration, we choose the central 
pressure $p_c$ and cooling rate $dS/dt$ (or equivalently cooling time 
$t_S=S/\left|dS/dt\right|$). For the inwards integration, we start at a 
pressure of 10\,bars and set the temperature there to be $T_{\rm iso}$. We then 
integrate inwards, choosing the luminosity $L$ and radius $R$. A 
multi-dimensional Newton-Raphson method is used to find the correct choices of 
($p_c,\,t_S,\,R,\,L$) that result in $m,\,r,\,T$ and $L$ agreeing to within 
1\% at the matching pressure.

As an example, Figure~\ref{f:structure} compares the entropy and
temperature profiles for models with an isothermal and non-isothermal
radiative zone. In the isothermal case, we choose
$T_c=3\times10^4\ {\rm K}$, $p_c=2\times\,10^{7}\ {\rm bars}$ and 
$T_{\rm iso}=1500\ {\rm K}$, which gives a $M=0.96\ M_J$, $R=1.25\ R_J$ planet 
with core entropy $S=7.98$ and convective zone boundary 
$p_{\rm conv}=62.76\ {\rm bars}$. The luminosity from the interior is 
$L=1.28\times\,10^{26}\ {\rm erg\ s^{-1}}$, giving $t_S=5.78\ {\rm Gyr}$. With 
a radiative zone, we obtain the same mass, radius and entropy with a 
convective zone boundary $p_{\rm conv}=131.7\ {\rm bars}$. The luminosity from 
the interior is $L=7.7\times 10^{25} \ \mathrm{erg\ s^{-1}}$, and 
$t_S=10.5\ {\rm Gyr}$ to cool.

In each case, the magnetic field structure in the planet interior is obtained 
by solving equation (\ref{eq:g}) using the conductivity profile in the planet 
(shown in Figure~\ref{f:sigma}), and then the ohmic heating profile is 
determined. For an induced magnetic field $B_{\phi 0}=10\,{\rm G}$ at the 
bottom of the wind zone ($p=10\ {\rm bars}$), we find 
$P_{\rm ohm}(p\geqslant{p_{\rm conv}})=8\times\,10^{23}\ \mathrm{erg\ s^{-1}}$ 
for the isothermal model and
$P_{\rm ohm}(p\geqslant{p_{\rm conv}})=6.8\times\,10^{22}\ \mathrm{erg\ s^{-1}}$ 
in the non-isothermal case. While the cooling time for the planet changes by a 
factor of two between the two models, the ohmic power changes by more than an 
order of magnitude. Therefore, it is crucial to locate the convective boundary 
accurately when calculating the ohmic power inside the convection zone.

In Figure~\ref{f:solution}, we compare the ohmic power calculated in this way, 
which includes the correct radial distribution of current, with the ohmic power 
calculated by assuming a constant radial current, independent of depth. The 
agreement is excellent (within a factor of 2) except at the highest pressures in the central regions 
of the planet. 

\begin{figure}
\epsscale{1.15}
\plotone{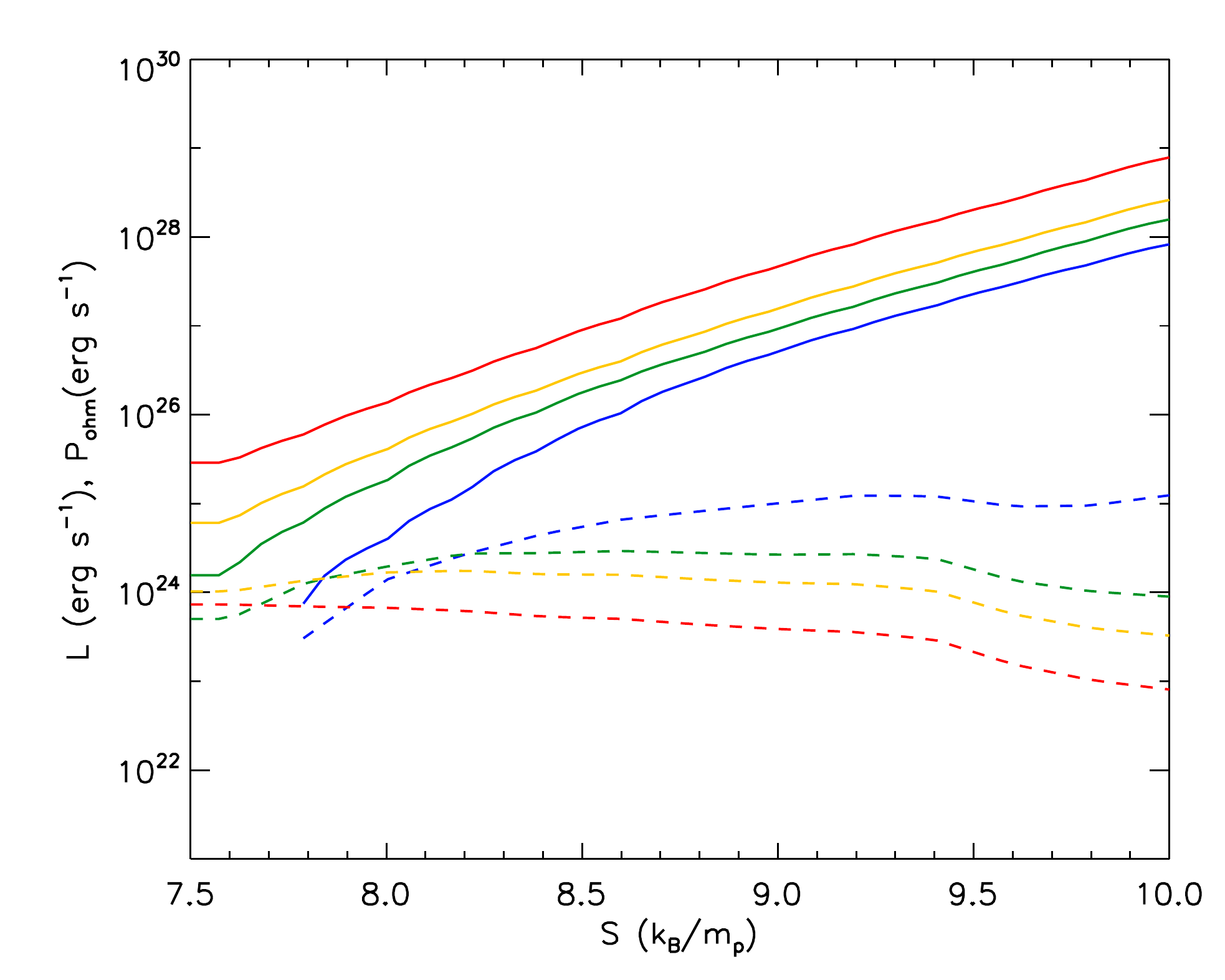}
\caption{
The luminosity $L$ (solid curves) and the total ohmic heating in the convection 
zone $P_{\rm ohm}$ (dashed curves) as a function of central entropy $S$. The 
red, yellow, green and blue lines (from top to bottom for the solid
lines; inverse for the dashed lines) represent planets with different mass: 
$3.0\ M_J$, $1.0\ M_J$, $0.6\ M_J$, $0.3\ M_J$. 
At larger entropy, where ohmic heating is unimportant, $L\propto M$ at fixed 
$S$, whereas $P_{\rm ohm}$ decreases with increasing $M$. All the planet models are computed with $T_{\rm iso}=1750\ {\rm K}$ and $B_{\phi 0}=100\ {\rm G}$.
\label{f:lpohm}}
\end{figure}


\begin{figure}
\epsscale{1.2}
\plotone{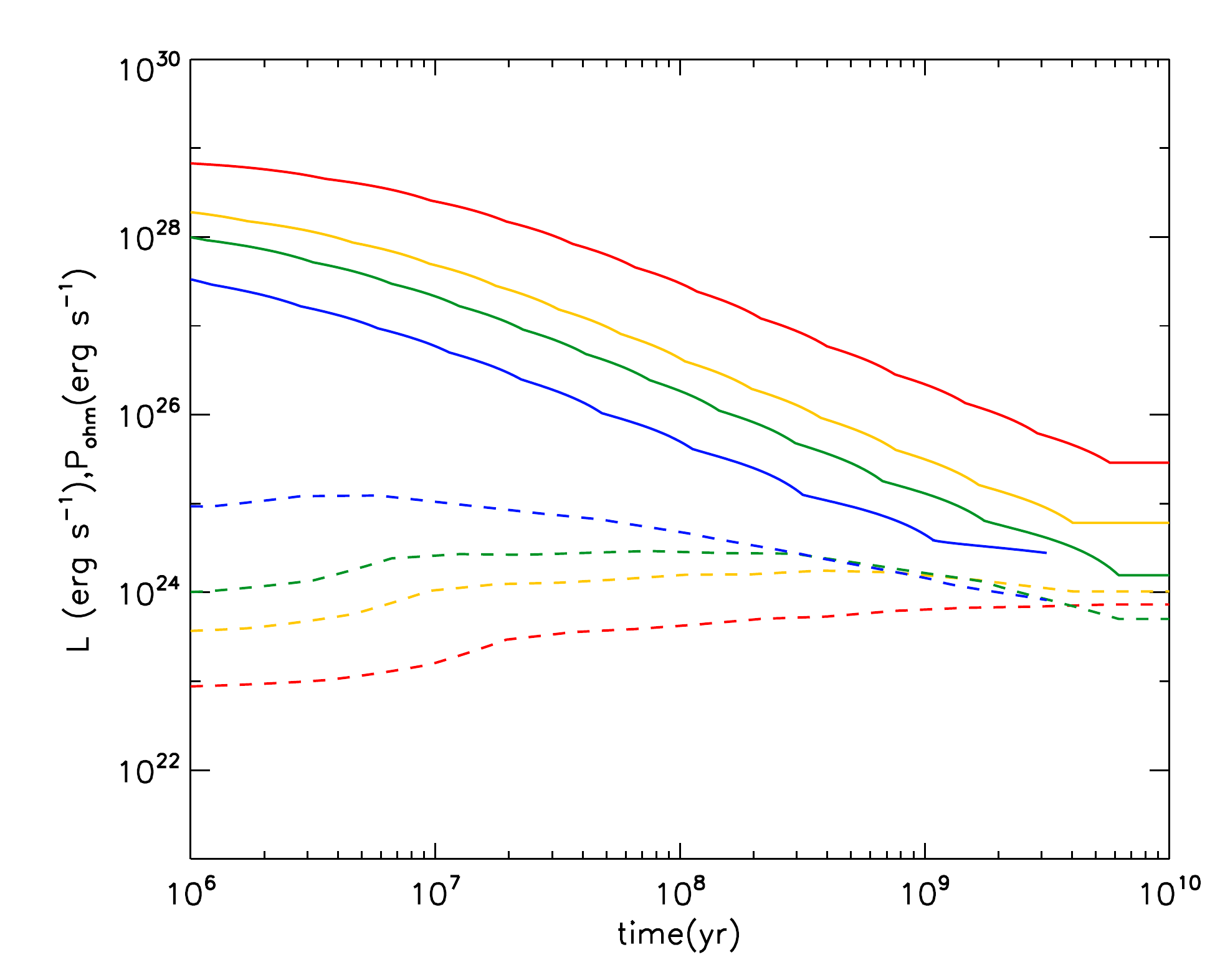}
\caption{
The time history of planet luminosity (solid curves) and ohmic heating in the 
convection zone $P_{\rm ohm}$ (dashed curves) for $M=0.3\ M_J$ (blue), 
$0.6\ M_J$ (green), $1\ M_J$ (yellow) and $3\ M_J$ (red curves) 
(same configuration with Figure \ref{f:lpohm}). The 
luminosity decreases with time because of cooling, while the ohmic heating 
either increases or decreases slowly depending on $M$. At late times, when 
ohmic heating in the radiative layer becomes important, $P_{\rm ohm}$ 
decreases because the convective boundary moves inwards. When $P_{\rm ohm}$ 
becomes comparable to $L$, the cooling and contraction of the planet is halted. 
All the models are calculated with $T_{\rm iso}=1750\ K$ and 
$B_{\phi\ 0}=100\ {\rm G}$. 
\label{f:luminosity}
}
\end{figure}

\begin{figure}
\epsscale{1.2}	
\plotone{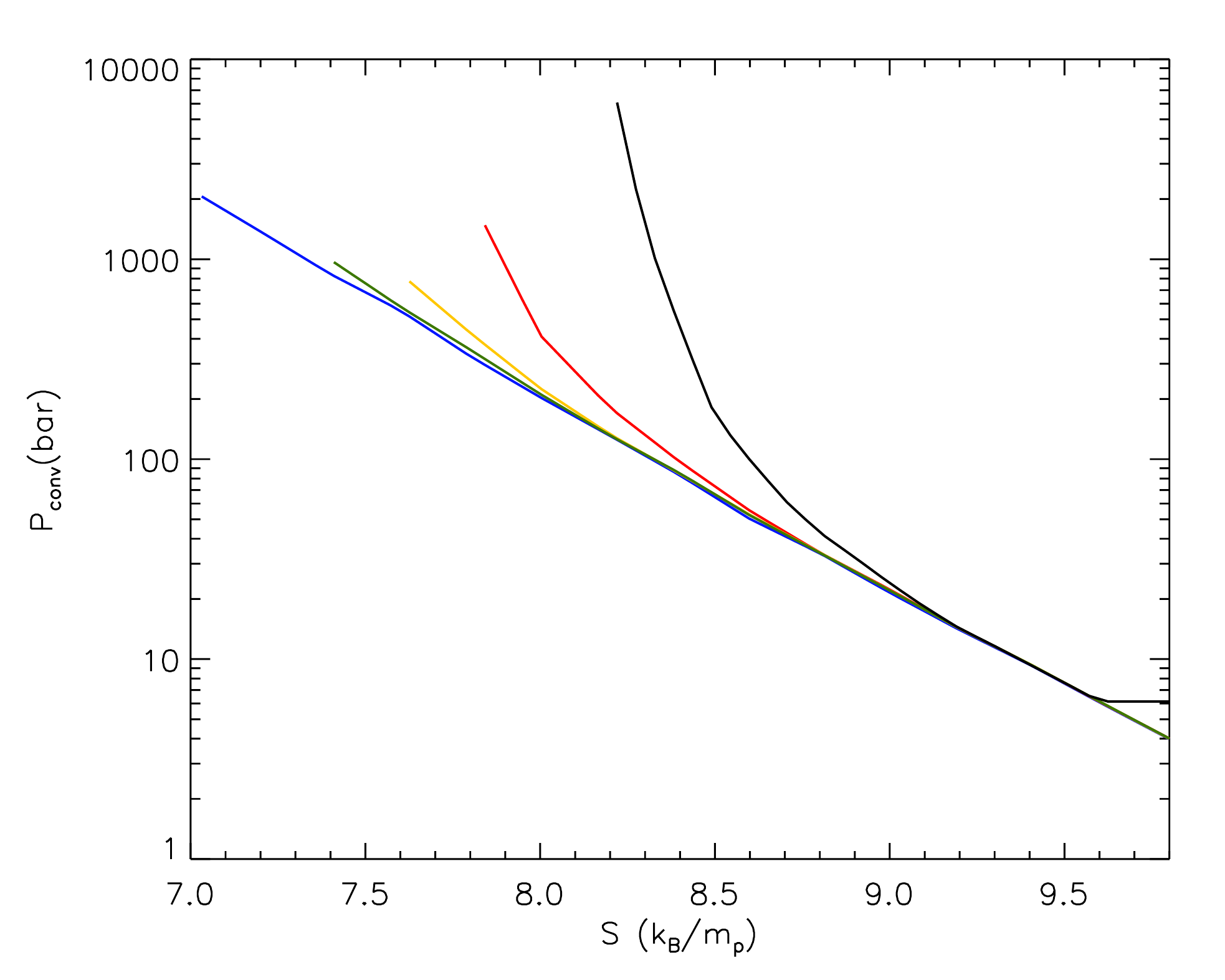}
\caption{
Position of the radiative--convective boundary as a function of internal 
entropy $S$ for a $1\ M_J$ planet with $T_{\rm iso}=1750\ {\rm K}$. Blue, 
green, yellow, red, black lines are for 
$B_{\phi 0}=$0, 10, 30, 100, 300, 1000 ${\rm G}$. At a given entropy, a larger 
$B_{\phi 0}$ results in more ohmic heating in the radiative zone, moving the 
convective boundary to a higher pressure. 
\label{f:pconv}
}
\end{figure}

\subsection{Planet models with feedback from ohmic heating}

The fact that the ohmic heating per unit mass rises rapidly to lower densities 
(Fig.~\ref{f:solution}) suggests that the heating in the regions lying between 
the wind zone and the convection zone boundary will be larger than the heating 
in the convective interior. We include the ohmic heating in the radiative 
layer by allowing $L$ to vary throughout the radiative zone, with
\begin{equation}
{dL\over dr}={J^2\over \sigma}.
\label{eq:feedback}
\end{equation}
We do not include ohmic heating at pressures less than 10 bars. Instead, we 
specify the temperature $T_{\rm iso}$ at $p=10$ bars and integrate inwards. 
Of course, there could be significant ohmic heating within the wind zone at 
$p<10$ bars, but we absorb this into the boundary condition. Note that this 
means that early in the lifetime of the planet, when the entropy is large 
enough that $p_{\rm conv}<10$ bars, the models here revert back to our 
previous models with no feedback. However, at those early times, ohmic heating 
is generally not yet important. Also note that in the models without feedback, 
the strength of the induced field $B_\phi$ does not influence the internal 
structure of the planet, whereas here a larger $B_\phi$ results in more 
heating in the radiative layer which can push the convective boundary deeper.

In Table \ref{t:models}, we compare the models with feedback to our earlier 
models without feedback. The internal structures are shown in 
Figure~\ref{f:structure}. The planet radius does not vary much between 
different models. The biggest difference is in the position of convective zone 
boundaries (marked by black vertical bars in Fig.~\ref{f:structure}), which 
results in a difference in the cooling luminosity and the ohmic heating both 
in the convective zone and atmosphere. Note that this means that the cooling 
history of a planet using these three approaches would be different, 
especially the time at which ohmic heating begins to become important for 
evolution. We use this feedback model for all the calculation carried on below. 

\subsection{Ohmic power as a function of entropy}

The luminosity and ohmic power is shown as a function of entropy in 
Figure~\ref{f:lpohm}. As noted in particular by \cite{Arras06}, equation 
(\ref{eq:lum}) shows that $L\propto M$ at fixed entropy, and we see that 
scaling in Figure~\ref{f:lpohm}. On the other hand, the ohmic power decreases 
with increasing $M$, as discussed in \S 2.2. We find that the decrease in 
$P_{\rm ohm}$ is well described by $P_{\rm ohm}\propto R^{2.4}/M$, which has a 
shallower dependence on $R$ than in equation (\ref{eq:pohmtotal}) because of 
the dependence of the conductivity term on mass which compensates the $R^4$ 
term. In our feedback model, for a lower mass planet the higher atmospheric 
ohmic heating pushes the convective zone boundary slightly deeper, resulting 
in a higher conductivity at the top of the convection zone. Overall, lower 
mass planets generally have higher ohmic power deposited in the convection 
zone. 

Combined with the mass-luminosity dependence, the decrease of ohmic power with 
mass means that ohmic heating becomes important for lower mass planets at a 
much higher entropy than for more massive planets. The value of entropy at 
which ohmic heating becomes important depends on the boundary induced field 
$B_{\phi 0}$. Turning this around, for an observed planet with measured 
radius and mass, we can infer the entropy and therefore derive a limit on the 
wind zone $B_{\phi 0}$ required for ohmic heating to be providing a 
significant part of the luminosity in that object. We carry out this procedure 
in \S \ref{observation}, but first describe our calculations of the 
time-evolution of planets with ohmic heating.


\section{Time Dependent Evolution of Planet Structure}
\label{evolution}

Having computed the luminosity at the radiative--convective boundary for a 
large grid of models with different $M$, $S$, $T_{\rm iso}$ and
$B_{\phi 0}$, the evolution in time of a planet with fixed mass $M$, 
$B_{\phi 0}$ and $T_{\rm iso}$ then involves stepping in entropy using 
equation (\ref{eq:dSdtohm}). When calculating the ohmic 
power, we assume constant current $J$ with depth, which as shown earlier is a 
good approximation. We have checked that for $B_{\phi 0}=0$, our cooling 
models compare well with the earlier results of \cite{Burrows1997} and 
\cite{Baraffe2003} (see \citealt{Marleau2012}). We reproduce their cooling 
curves to within 30\% in luminosity, and predict radii that are 
$0.05$-$0.08\ R_J$ larger than in those cooling sequences. 

\begin{figure}
\epsscale{1.2}	
\plotone{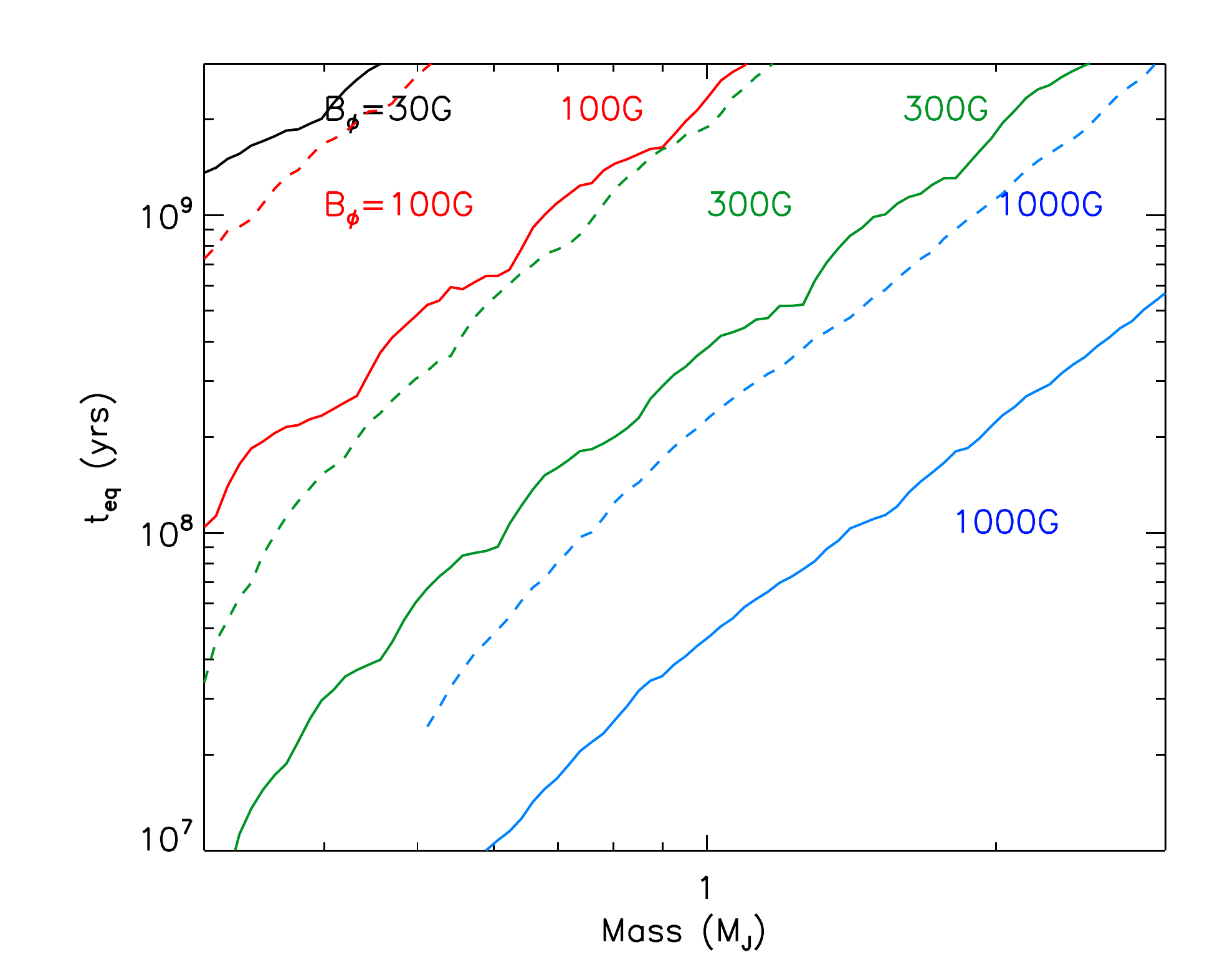}
\caption{
The age of the planet when ohmic heating becomes important 
($P_{\rm ohm}>0.1 L$) versus planet mass, for $B_{\phi 0}=30\ {\rm G}$ (black), 
$100\ {\rm G}$ (red), $300\ {\rm G}$ (green), $1000\ {\rm G}$ (blue curves), 
and for two temperatures $T_{\rm iso}=1750\ {\rm K}$ (solid curves) and 
$2250\ {\rm K}$ (dashed curves, corresponding to the lower panel of labels).
\label{f:age}
}
\end{figure}

\begin{figure}
\epsscale{1.15}	
\plotone{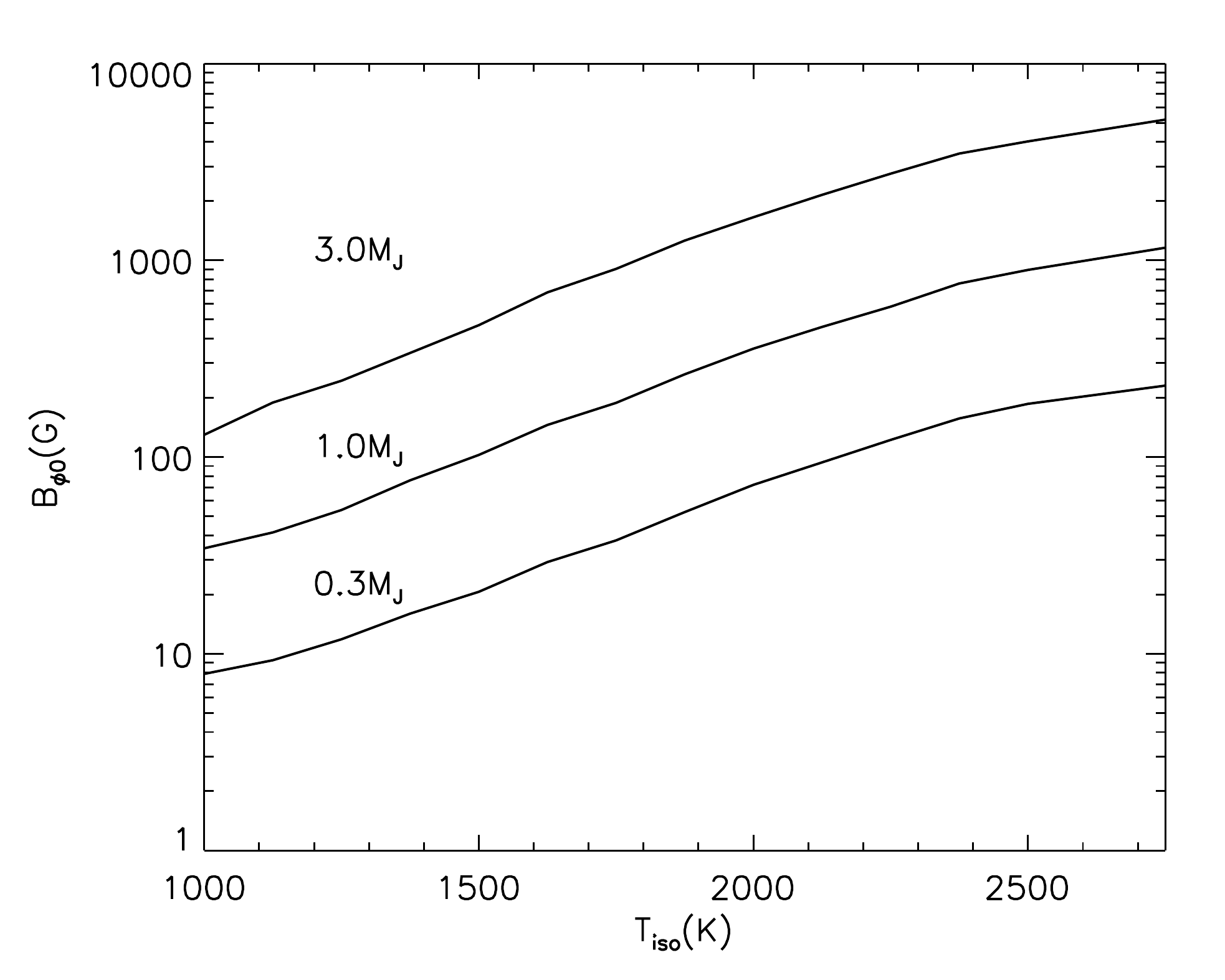}
\caption{
The value of $B_{\phi 0}$ required for ohmic heating to become important at an 
age of $3\ {\rm Gyr}$. From bottom to top, $M=$0.3, 1, 3.0 $M_J$.
\label{f:Biso}
}
\end{figure}

\begin{figure}
\epsscale{1.2}	
\plotone{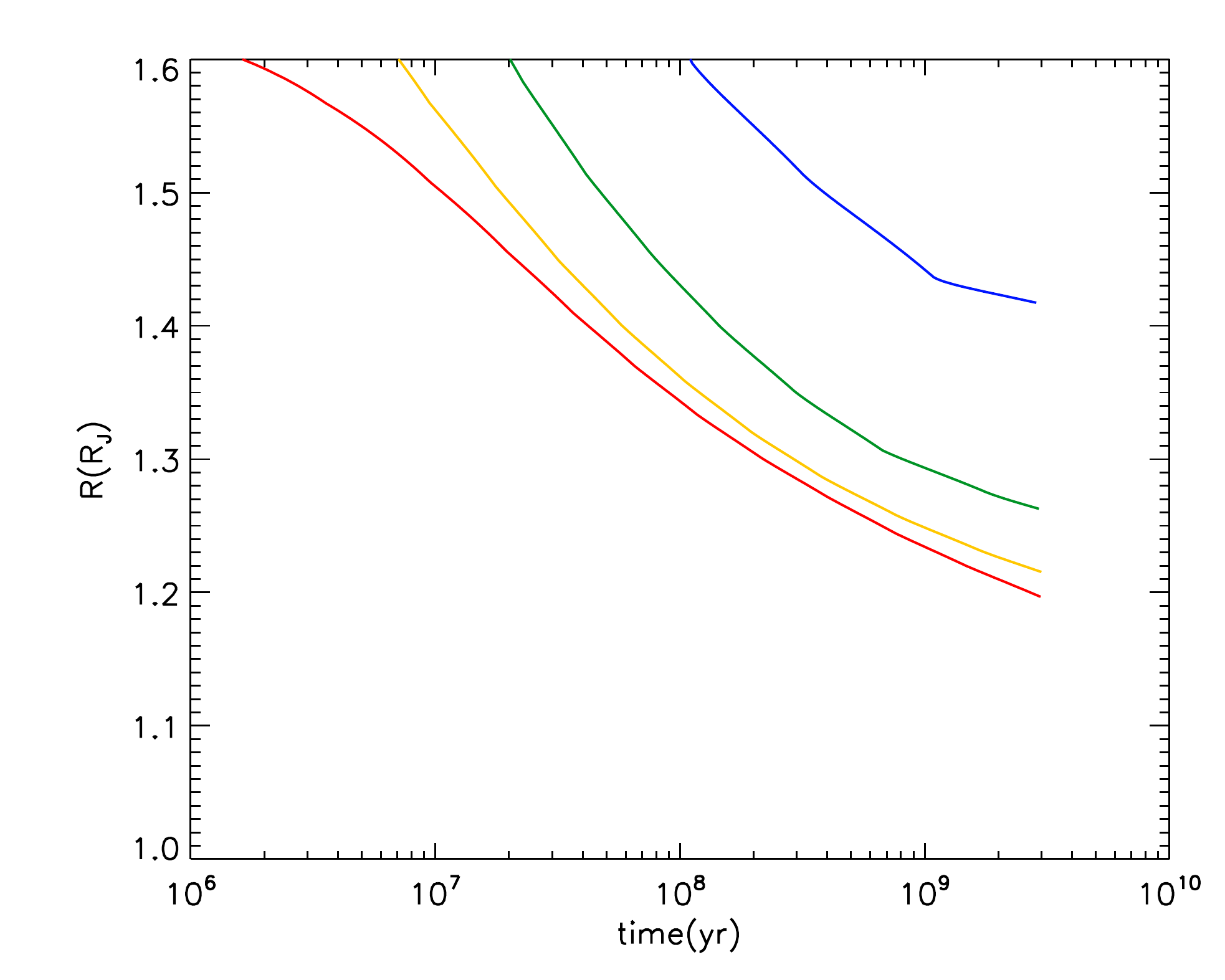}
\caption{
Time history of planet radius for (top to bottom) $M=$0.3, 0.6, 1 and $3\ M_J$. 
All the models are calculated with $T_{\rm iso}=1750\ {\rm K}$ and
$B_{\phi 0}=100\ {\rm G}$. 
\label{f:cooling2}
}
\end{figure}

By integrating in time, we compute the time history of the planet luminosity 
and ohmic power, shown in Figure~\ref{f:luminosity}. As the planet cools, the 
convection zone ohmic power increases or decreases slightly depending on mass, 
but always changes more slowly than $L$, so that ohmic power eventually 
becomes comparable to the cooling luminosity. At the same time, as the ohmic 
heating in the upper atmosphere (which is about an order of magnitude larger 
than convection zone heating) starts to affect the planet structure, the 
convective zone boundary shrinks inwards. The resulting decrease of both 
cooling luminosity and convection zone ohmic heating result in a rapid 
increase in cooling time, so that we can view the evolution afterwards as a 
quasi-steady state. For higher atmospheric ohmic heating, this effect happens 
at higher entropy, thus the steady radius of planet will be larger. We compare 
the evolutionary tracks of the radiative/convection zone boundary of a 1\,$M_J$ 
planet with different strengths of induced field in Figure~\ref{f:pconv}. As 
we increase the amount of ohmic heating, the convection zone boundary deviates 
from the no heating path at a higher entropy.

In Figure~\ref{f:age}, we show the age of a planet with a particular mass when 
$P_{{\rm ohm},c}=0.1L_{\rm conv}$, at which point ohmic heating starts to 
become significant and the planet contraction slows. In general, the 
atmospheric heating is an order of magnitude larger, and so comparable to the 
cooling luminosity at this age. In Figure~\ref{f:age}, we report that the 
result is sensitive to the strength of $B_{\phi 0}$.  With 
$T_{\rm iso}=1750\ {\rm K}$, for $B_{\phi 0}$ equals $30\ {\rm G}$, a 
$0.3\ M_J$ hot jupiter can reach steady state in $~1\ {\rm Gyr}$, while a 
$1\ M_J$ requires $100\ {\rm G}$ to reach steady state at a similar age. 
Higher $T_{\rm iso}$ (dashed line in Figure~\ref{f:age}) does not help the 
planet reach a steady-state radius faster, but on the contrary, it requires a 
larger induced field to achieve the same result. 

Figure~\ref{f:Biso} shows the $B_{\phi 0}$ required to halt contraction 
within 3 Gyr as a function of $T_{\rm iso}$. We see that a hotter planet 
requires a stronger induced field to obtain a significant level of ohmic 
heating. This is because the interior ohmic heating is closely related to the 
conductivity at the bottom of the wind zone; the conductivity increases with 
temperature, reducing the ohmic heating at fixed induced field. But we should 
also point out that for a hotter planet there is a higher chance to obtain a 
stronger induced field due to the stronger wind in the atmosphere. So this 
result does not necessarily imply that it is more difficult to make ohmic 
heating important in hotter planets. 

We plot the time evolution of the planet radius for planets with 
$B_{\phi 0}=100\ {\rm G}$ and $T_{\rm iso}=1750\,{\rm K}$ and different planet 
masses in Figure~\ref{f:cooling2}. In the absence of stellar ages, we shall 
take a typical age of 3 Gyr, and take the radius at $3\ {\rm Gyr}$ as the 
present day radius. For the lowest mass planet, $0.3 M_J$, the effect on the 
evolution of the planet radius is significant, and the planet stops cooling 
around $1\ {\rm Gyr}$ (with cooling time longer than $10\ {\rm Gyr}$) and 
thereafter maintains a large radius. However, the heating is not as effective 
at higher masses. For example, HD~209458b has a observed radius of $1.35\ R_J$ 
with mass $0.7\ M_J$, while we can only obtain a radius of $1.25\ R_J$ for 
$B_{\phi 0}=100$ G. This is because the power we introduced into the planet 
interior is far smaller than the received stellar luminosity. In the case of 
our standard model, the irradiation luminosity from the host star is 
$10^{29}\ {\rm erg\ s^{-1}}$, and the heating in the interior is only one 
$0.01\%$ of it, $10^{26}\ {\rm erg\ s^{-1}}$. To go further, we must 
understand what values of $B_{\phi 0}$ might be expected as a function of 
$T_{\rm eq}$, and we turn to this in the next section.


\section{Evolution including wind zone model and comparison to observations}
\label{observation}

In \S\ref{evolution}, we calculated the time-evolution of cooling gas giants 
assuming that $T_{\rm iso}$ and $B_{\phi 0}$ are independent parameters. 
In reality, they are coupled by the dynamics in the wind zone, since the 
atmospheric flow, in response to the irradiation, determines both the magnetic 
field in the layer and the temperature at depth (the values of $T_{\rm iso}$ 
and $B_{\phi 0}$ are specified at $p=10\ {\rm bars}$). In this section, we 
implement the scalings for the wind zone dynamics proposed by 
\cite{Menou2012} (\S \ref{sec:menou}) and then compare our results to observed 
systems (\S \ref{sec:obs}).

\subsection{Dynamics of the wind zone and the relation between $T_{\rm iso}$ and $B_{\phi 0}$}
\label{sec:menou}

Both \cite{B&S2011} and \cite{Menou2012} write down simplified models for the 
wind zone dynamics including the effects of magnetic drag. In both cases, 
following \cite{Perna2010a} the magnetic drag force is assumed to be 
$\vec{J}\times\vec{B}/c$ per unit volume, with the current $\vec{J}$ set by a 
balance between the shearing of the magnetic field by the fluid and ohmic 
diffusion of magnetic field lines against the fluid motion, 
$\vec{J}=\sigma\vec{v}\times\vec{B}/c$ (\S \ref{Con}). However, the dynamical 
balance in the two models is quite different. \cite{Menou2012} writes the 
force balance for the equatorial flow as (see also \citealt{Showman2010})
\begin{equation}
\label{eq:vphi}
0=-\frac{v_{\phi}^2}{R_P}+\frac{\mathcal{R}{\Delta}T_{\rm horiz}\Delta \ln{p}}{R_P}-\frac{v_{\phi}B_r^2}{4\pi\rho\eta}.
\end{equation}
The first two terms represent a balance between the advective term and the 
horizontal driving from the day-night temperature difference 
$\Delta T_{\rm horiz}$. This balance is thermal wind-like in that the 
horizontal pressure gradients require a vertical gradient in the fluid 
velocity $v_\phi$ over a vertical pressure scale $\Delta\ln p$. The final term 
represents the magnetic drag force, again integrated over a vertical scale 
$\Delta \ln p$. \cite{B&S2011} on the other hand consider the meridional 
circulation induced by magnetic drag on the azimuthal flow, so that for 
example the latitudinal force balance is $f v_y=v_\phi/\tau_L$ where 
$f=2\Omega\sin\theta$ is the Coriolis parameter and $\tau_L$ the magnetic drag 
timescale. Their solution represents a thermal wind balance involving the 
equator-pole temperature gradient, modified by magnetic drag. 

In both cases, magnetic drag limits the fluid velocity at high temperatures, 
where the large degree of ionization and therefore large electrical 
conductivity results in strong coupling of the fluid and magnetic field. 
Balancing the second and third terms in equation (\ref{eq:vphi}) gives 
$v_\phi\propto \eta$ when magnetic drag dominates, and therefore the magnetic 
Reynolds number $R_M=v_\phi H/\eta$ becomes almost constant, varying only 
slowly with temperature. Similarly, equation (16) of \cite{B&S2011} has two 
possible limits, either $v_\phi\propto \eta$ when the lateral temperature 
gradient is large, in which case $R_M$ becomes almost constant at large 
$T_{\rm iso}$, or $v_\phi\propto \eta^2$ when the drag time scale is 
comparable to the rotation period, while the lateral gradient of temperature 
is still small, in which case $R_M\propto \eta$ declines rapidly at large 
$T_{\rm iso}$.

\begin{figure*}
\plottwo{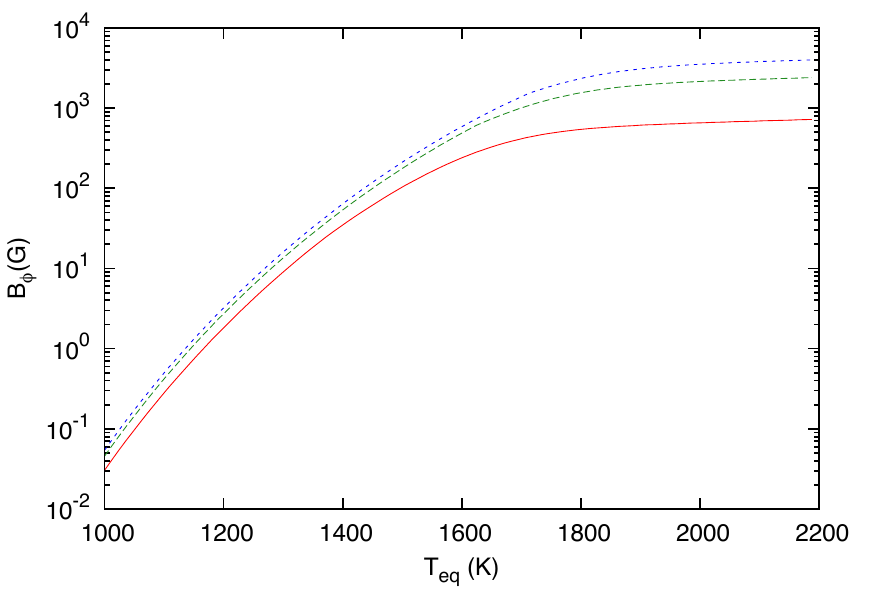}{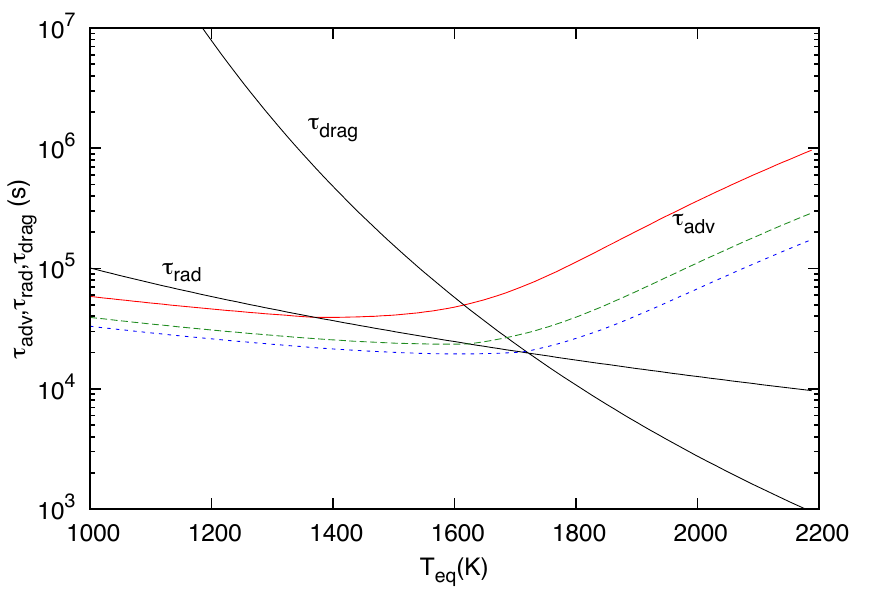}
\caption{
The induced field $B_\phi$ (left panel) and timescales (right 
panel) in the wind zone as a function of $T_{\rm eq}$. In the left panel, the 
solid, dashed and dotted curves are for wind zone thickness 
$\Delta \ln p=0.9, 3$ and $5$, and we take $B_r=10\ {\rm G}$. In the right 
panel, we show the advection, radiative and drag timescales 
$\tau_{\rm adv}$, $\tau_{\rm rad}$ and $\tau_{\rm drag}$. The advective 
timescale is shown for $\Delta \ln p=0.9$ (solid), $3$ (dashed) and $5$ 
(dotted curves) (the radiative and drag timescales are independent of 
$\Delta\ln p$).
\label{f:wz2}
}
\end{figure*}

A dynamical model including both day-night driving and meridional circulation 
with magnetic drag is not yet available. For our purposes, we have implemented 
the model of \cite{Menou2012} as described by equation (\ref{eq:vphi}), with 
the day-night temperature difference given by
\begin{equation}\label{eq:Thoriz}
{\Delta}T_{\rm horiz}=\cases{
\frac{T_{\rm day}}{2}(\frac{\tau_{\rm adv}}{\tau_{\rm rad}}) & $\tau_{\rm adv}<\tau_{\rm rad}$\cr\frac{T_{\rm day}}{2} & $\tau_{\rm adv}>\tau_{\rm rad}$\cr}
\end{equation}
In equation (\ref{eq:Thoriz}), $T_{\rm day}$ is the dayside averaged 
temperature considering a dilution factor of 0.5, 
$T_{\rm irr}^4=2T_{\rm day}^4=4T_{\rm eq}^4$, and the advective and radiative 
timescales are
\begin{equation}
\tau_{\rm adv}=\frac{R_P}{v_{\phi}}
\end{equation}
\begin{equation}\label{eq:trad}
\tau_{\rm rad}=\frac{C_pp}{g{\sigma_{SB}}T_{\rm day}^3}.
\end{equation}
Note that these timescales are evaluated at the outermost pressure, which 
following \cite{Menou2012} is taken to be $60\ {\rm mbars}$, the estimated 
location of the thermal photosphere. This is the reason for adopting the 
thermal timescale appropriate for an optically thin region, so that 
$\tau_{\rm rad}\propto p$ in equation (\ref{eq:trad}); in deeper, optically 
thick layers, $\tau_{\rm rad}$ has an extra factor of the optical depth 
$\tau$, leading to $\tau_{\rm rad}\propto p^2$ (e.g.~Fig.~3 of 
\citealt{Showman2008}). The magnetic drag term is also evaluated at 
$p=60$ mbars; this term is integrated over height, but since $\sigma$ 
decreases with increasing pressure (for an isothermal layer), the dominant 
contribution to the integral is from the lower limit on pressure, and so 
$\eta$ and $\rho$ are evaluated there. This is an important difference from 
\cite{B&S2011}, who evaluated their magnetic drag timescale at 
$p=10\ {\rm bars}$, which gives a drag time an order of magnitude longer than 
we find here. Based on that estimate, \cite{B&S2011} concluded that the drag 
timescale was always much longer than a rotation period.

We solve equation (\ref{eq:vphi}) for $v_\phi$ as a function of $T_{\rm eq}$, 
and find the corresponding value of the induced field $B_\phi$ from equation 
(\ref{eq:RM}) for different values of the dipole field $B_{\rm dip}$. For 
$\Delta\ln p=0.9$, we reproduce the results of \citet{Menou2012} (see his 
Fig.~1), but we also consider larger values of $\Delta\ln p$. \citet{Menou2012} 
models the weather layer with a modest vertical extension around 1 pressure 
scale height. We also solve the equation with $\Delta\ln{p}=3$ for typical 
values in hot jupiter atmosphere as reported by \citet{Showman2010}, and 
$\Delta\ln{p}=5$ for a wind zone extending to $p\sim 10$\ bars, for comparison 
with \cite{B&S2010} and \cite{B&S2011}. The effect of varying 
$\Delta \ln p$ on $B_\phi$ is shown in the left panel of Figure~\ref{f:wz2}. 
For numerical convenience, we fit the $B_\phi$--$T_{\rm eq}$ relation with the 
following:
\begin{equation}
\frac{1}{B_{\phi}(T_{\rm eq})}=\frac{1}{B_{\rm adv}}+\frac{1}{B_{\rm drag}},
\end{equation}
where
\begin{eqnarray}\label{eq:Badv}
B_{\rm adv}&=&2.8\times 10^6\ {\rm G}\  T_{\rm eq}
\exp\left(-{2.53\times 10^4\over T_{\rm eq}}\right)\nonumber\\&&
\left({\Delta\ln{p}\over 3}\right)^{1/2}\left({B_r\over 10\ {\rm G}}\right)
\end{eqnarray}
and
\begin{equation}\label{eq:Bdrag}
B_{\rm drag}=1125\ {\rm G}\ \left({T_{\rm eq}\over 1000\ {\rm K}}\right)\left({\Delta\ln p\over 3}\right)\left({B_r\over 10\ {\rm G}}\right)^{-1}.
\end{equation}
This reproduces $B_\phi$ to within $\approx 10$\% for $T_{\rm eq}$ in the range $1100$ to $2200$\ K.

The transition to the regime where $R_M$ is approximately constant occurs when 
$\tau_{\rm drag}$ exceeds $\tau_{\rm adv}$, where the magnetic drag timescale 
is $\tau_{\rm drag}=4\pi\rho\eta/B_r^2=\eta/v_A^2$, where $v_A$ is the Alfven 
speed, and again the timescale is evaluated at the top of the wind zone 
($p=60$ mbars here). These timescales are plotted as a function of $T_{\rm eq}$ 
in Figure~\ref{f:wz2}. Changing the wind zone thickness from $\Delta \ln p=0.9$ 
to $\Delta \ln p=5$ moves the transition temperature from 
$T_{\rm eq}\approx 1400$\ K to 1700\ K. 

To use this value of $B_\phi$ as a boundary condition for our evolutionary 
models, we must relate the temperature $T_{\rm eq}$ at low pressure to the 
temperature $T_{\rm iso}$ at $p=10$\ bars. This relation depends on the details 
of energy transport in the wind zone, including the effects of ohmic heating 
and needs to be studied further. Here, we adopt the atmospheric temperature 
profile from \citet{Guillot2010}, and keep in mind the uncertainty in the 
relation between $T_{\rm eq}$ and $T_{\rm iso}$ when interpreting our results 
below. The relation from \citet{Guillot2010} is (see his eq.~[29])
\begin{equation}\label{eq:guillot}
T^4=\frac{3T_{\rm irr}^4}{4}f\left[\frac{2}{3}+\frac{1}{\gamma\sqrt{3}}\right],
\end{equation}
where $\gamma$ is the ratio between visible and infra-red opacities, 
and $f=1/2$ for a dayside average or $f=1/4$ for an average over the whole 
surface. Choosing $\gamma=0.4$, as appropriate for a planet like 
HD~409658b (e.g.~see Fig.~1 of \citealt{Hubeny2003}) and a dayside average 
$f=0.5$, we obtain $T_{\rm iso}=0.94T_{\rm irr}=1.33T_{\rm eq}$. In the 
following section, we will use this relation to infer the appropriate value of 
$T_{\rm iso}$ from the $T_{\rm eq}$ of observed planets. We note here
that we don't have a good knowledge of what the $\gamma$ parameter would
be for most of the observed planets. While $\gamma$ parameter could vary
in a very large numerical range, the ratio between $T_{\rm iso}$ and
$T_{\rm eq}$ only changes within a factor of few
[$0.99(\gamma\to\inf)<(T_{\rm {iso}}/T_{\rm {eq}})<3.05(\gamma=0.01)$].     
Since the observed properties of planets gives $T_{eq}$ thus the boundary 
induced field, changing the value $T_{\rm iso}/T_{\rm eq}$ is equivalent to shift 
inside the plane $T_{\rm iso}-B_{\phi}$ given by Figure \ref{f:Biso}. 
Generally, a smaller $T_{\rm iso}/T_{\rm eq}$ is favored to inflate the planet with
the same $B_{\phi}$.


\begin{figure}
\epsscale{1.1}
\plotone{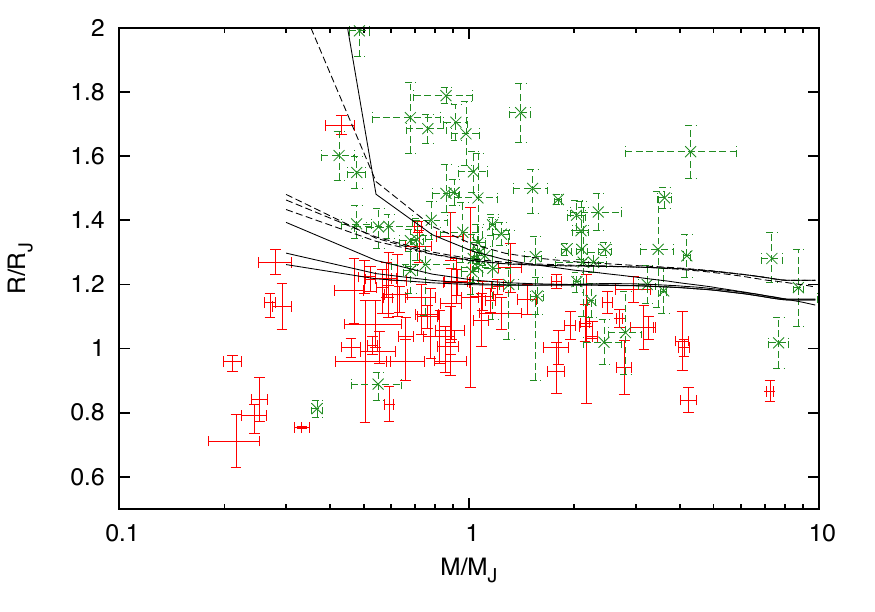}
\caption{
The predicted mass--radius relation at $3\ {\rm Gyr}$ for 
$B_{\phi 0}=$0, 30, 100 and $1000\ {\rm G}$ and $T_{\rm eq}=1316\ {\rm K}$ 
(solid curves) and $1692\ {\rm K}$ (dashed curves). The data points show 
observed transiting planets, divided into two temperature groups 
$T>1500\ {\rm K}$ (green points) and $T<1500\ {\rm K}$ (red points).
\label{f:MRB}
}
\end{figure}


\subsection{Comparison with Observed Hot Jupiters}
\label{sec:obs}

\begin{figure}
\epsscale{1.2}
\plotone{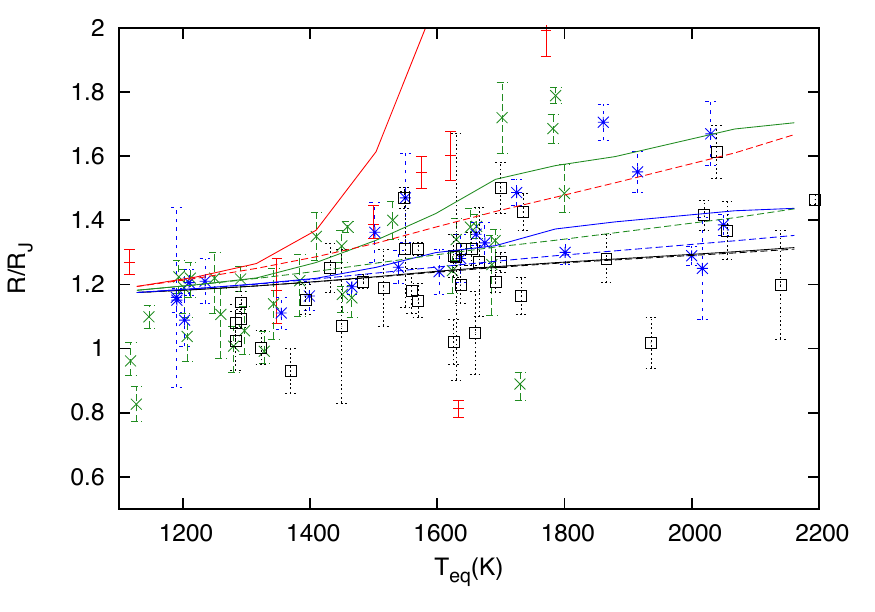}
\plotone{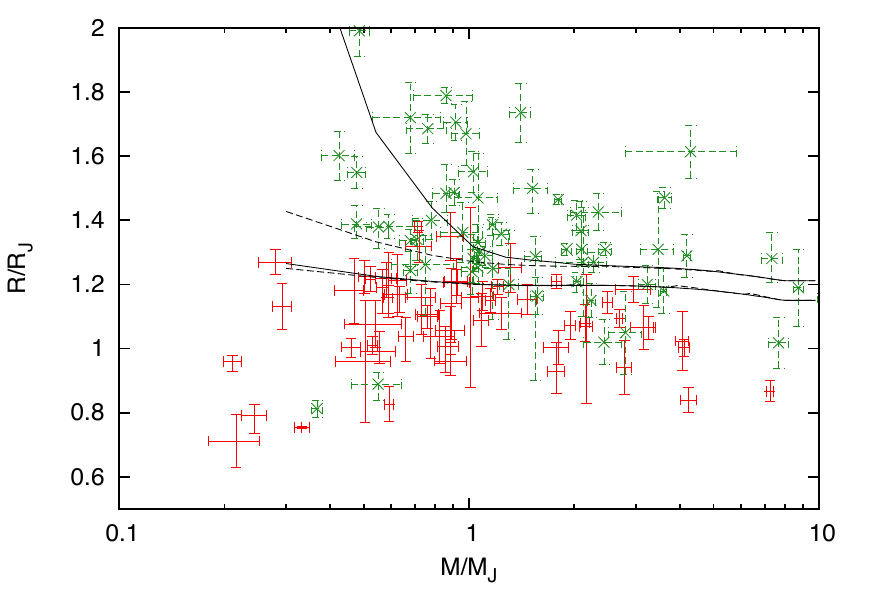}
\caption{
Comparison with observations using $B_{\phi 0}(T_{\rm eq})$
from the wind zone model. {\em Top panel:} Radius at 3 Gyr against
$T_{\rm eq}$ for $M=0.3$ (red), $0.6$ (green), $1.0$ (blue) and $3.0\
M_J$ (black). The data points are observed planets divided by mass:
$0.2\ M_J<M<0.5\ M_J$ (red points), $0.5\ M_J<M<0.9\ M_J$ (green
points), $0.9\ M_J<M<1.3\ M_J$ (blue points). {\em Bottom panel:}
Predicted mass--radius relation at $3\ {\rm Gyr}$ for 
$T_{\rm eq}=1316\ {\rm K}$ and $T=1682\ {\rm K}$ (bottom to top). In each case, 
the dashed curve shows the radius without ohmic heating; the solid curve with 
ohmic heating. The data has been divided by temperature: 
$T_{\rm eq}<1500\ {\rm K}$ (red points), $T>1500\ {\rm K}$ (green points). 
\label{f:observ}
}
\end{figure}

\begin{figure}
\epsscale{1.2}
\plotone{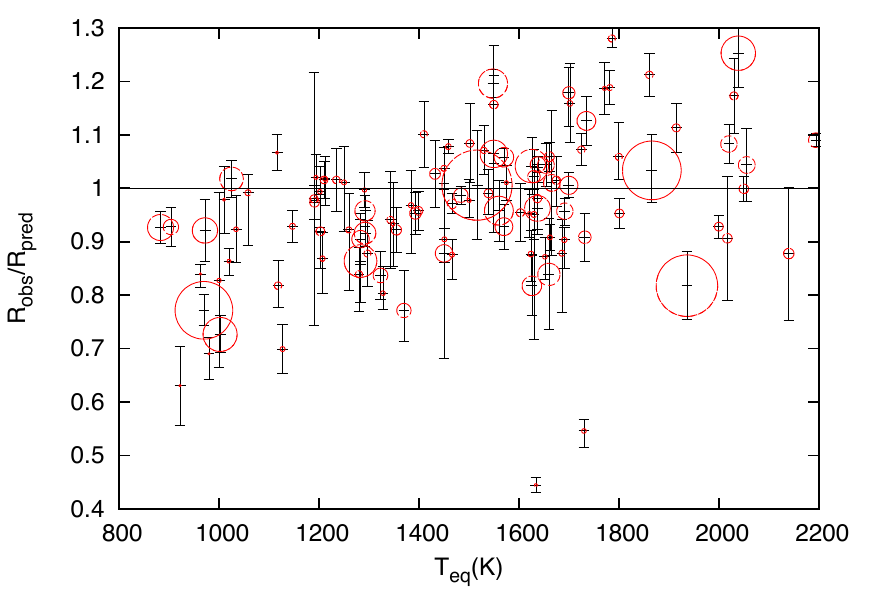}
\plotone{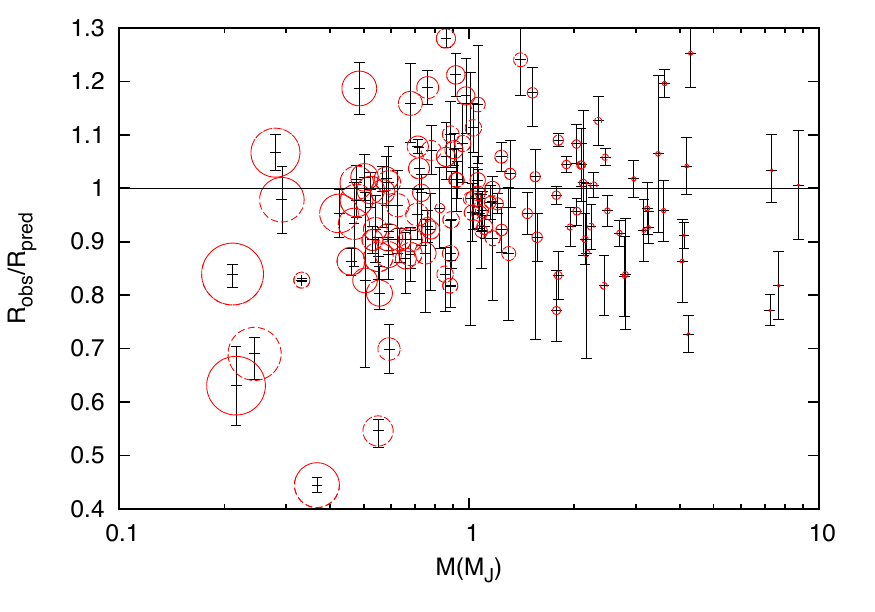}
\caption{
The ratio of observed planet radius $R_{\rm obs}$ and predicted radius 
$R_{\rm pred}$ (3 Gyr) for observed hot jupiters as a function of $T_{\rm eq}$ 
(upper panel) or $M$ (lower panel). In the upper panel, the size of the circle 
scales with planet mass; in the lower panel, the size of the circle scales 
with $T_{\rm eq}$.
\label{f:RPredict}
}
\end{figure}

In Figures \ref{f:MRB} to \ref{f:RPredict}, we compare our results with the 
observed properties of transiting planets taken from the TEPcat transiting 
planet catalog\footnote{http://www.astro.keele.ac.uk/$\sim$jkt/tepcat/}, 
which gives the planet mass, radius, and equilibrium temperature 
$T_{\rm eq}=T_{\star,{\rm eff}}\,(R_\star/2a)^{1/2}$ where 
$T_{\star,{\rm eff}}$ is the stellar effective temperature. As the ages of 
most stars are unknown or highly uncertain, we assume an age of 3 Gyr when 
comparing with the observed planets.

First, Figure~\ref{f:MRB} shows the effect of increasing $B_{\phi 0}$ at 
fixed $T_{\rm iso}$ on the planet radius. To help compare with the data, we 
divide the observed sample into two groups with either 
$T_{\rm eq}>1500\ {\rm K}$ (green points) or $<1500\ {\rm K}$ (red points) and 
show theoretical curves for either $T_{\rm eq}=1316\ {\rm K}$ or 
$1692\ {\rm K}$ (these two temperatures correspond to $T_{\rm iso}=1750$ and 
$2250\ {\rm K}$ respectively). We see that for the low $T_{\rm eq}$ group, an 
induced field of 10--100\,G can explain most of the observed radii, while the 
high $T_{\rm eq}$ planets need at least $B_{\phi 0}=1000\ {\rm G}$ to match 
the observed radii. It is clear that a higher induced magnetic field is needed 
to explain a given radius at higher equilibrium temperature. 

Next, we use the wind zone model described in \S \ref{sec:menou} to calculate 
$B_{\phi 0}$ as a function of $T_{\rm eq}$, assuming canonical values 
$B_{r}=10\ {\rm G}$ and $\Delta\ln{p}=3$. In the top panel of 
Figure~\ref{f:observ}, we show the radius as a function of $T_{\rm eq}$, 
with the colors representing three different bins in planet mass. There 
exists a clear correlation between the radius and $T_{\rm eq}$, both in the 
observations and the models. In addition, we see that the amount of inflation 
is also strongly dependent on the planet mass. Planets within the mass bin 
0.3--0.6 $M_J$ agree quite well with our ohmic heating model. However, ohmic 
heating clearly cannot explain planets with mass $\sim1\ M_J$ and large 
inflated radii $\gtrsim 1.4\ R_J$. Ohmic heating can help to increase the 
radius (for comparison the dashed line shows models with no ohmic heating), 
but not enough to match the observed value. This is a consequence of the 
increased power needed to maintain the radius of a massive planet at a 
particular value, as well as the reduced ohmic heating power at larger masses.

In the lower panel of Figure~\ref{f:observ}, we show the radius as a function 
of mass. As in Figure~\ref{f:MRB}, we divide the data into two temperature 
ranges and show the model results for two representative temperatures, now 
using the wind zone model to specify the value of $B_{\phi 0}$ for each 
temperature. The parameter region where ohmic heating has the largest effect 
is high temperature, low mass planets. The radii of the low temperature group 
($T_{\rm eq}<1500$ K) can almost all be explained without ohmic heating. For 
the high temperature group, ohmic heating can explain the observed radii of 
low mass planets, but most of the radii of the high temperature group lie well 
above the models, especially at large planet masses $\gtrsim 1\ M_J$.

In Figure~\ref{f:RPredict}, we show the ratio between observed and predicted 
radii $R_{\rm obs}/R_{\rm pred}$ against $T_{\rm eq}$ (upper panel) and 
against $M$ (lower panel) for each observed planet. In this case, we use the 
observed values of $M$ and $T_{\rm eq}$ to calculate the evolution of the 
planet, and, in the absence of stellar ages, we take $R_{\rm pred}$ to be the 
radius at 3 Gyr. In the upper panel, we see that for 
$T_{\rm eq}\gtrsim 1600\ {\rm K}$, there are many planets whose radii lie 
above the predicted values. The slow increase of $B_{\phi 0}$ with 
$T_{\rm iso}$ at large $T_{\rm iso}$ due to the magnetic drag term results in 
a much weaker dependence of $R_{\rm pred}$ on $T_{\rm eq}$ than observed, and 
most outliers lie at the highest temperatures. In the lower panel, we see that 
the majority of the unexplained objects ($R_{\rm obs}>R_{\rm pred}$) are at 
larger masses $M\gtrsim 0.7\ M_J$. We note that the choice of 
estimating the planet radius at $3\ {\rm Gyr}$ is not critical for the above 
picture. Constraining ourself within the time range of $1-5\ {\rm Gyr}$,
the predicted radius only varies within several percent.

\begin{figure*}
\epsscale{1.0}
\plotone{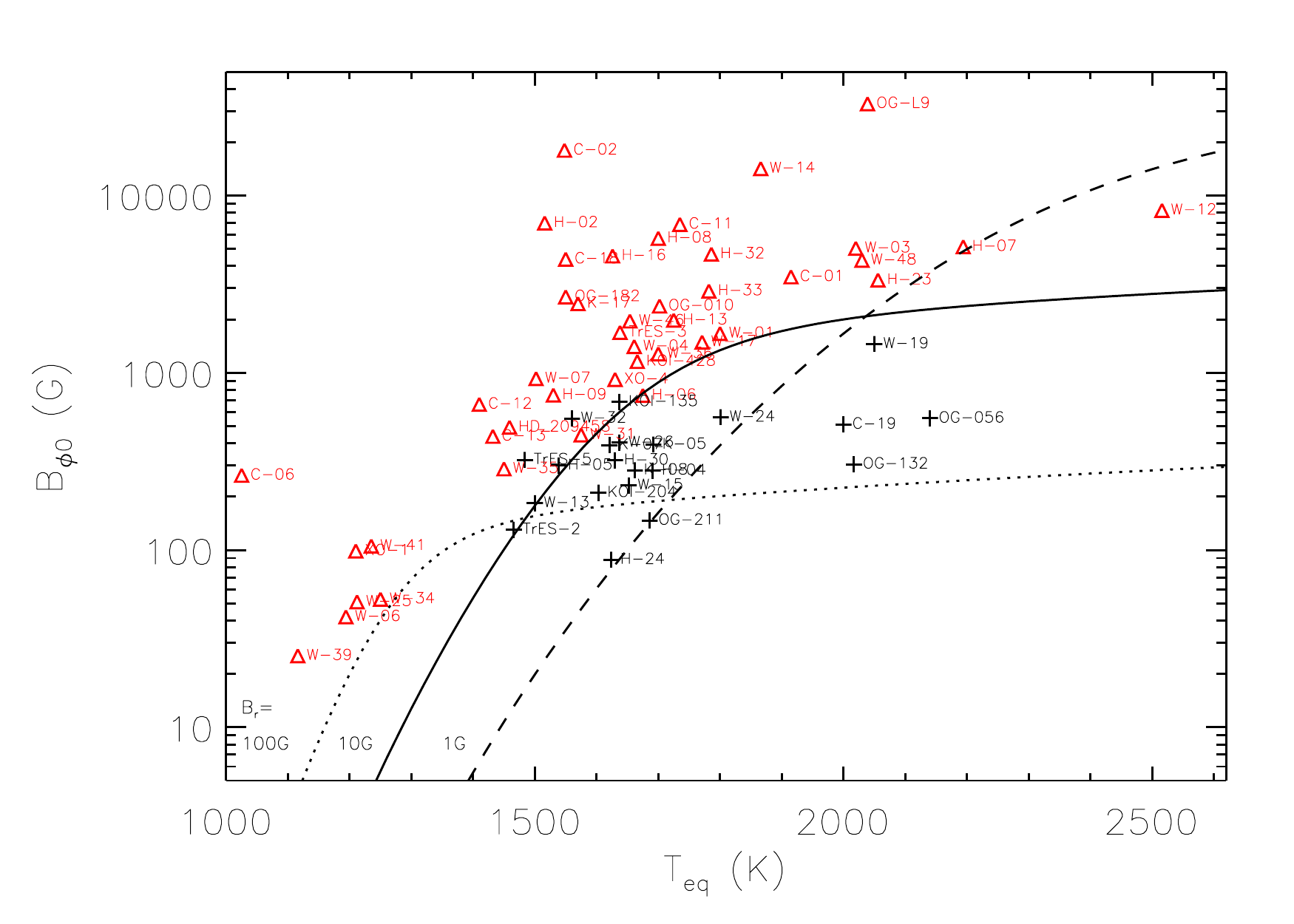}
\caption{
For each observed planet, we show the $B_{\phi 0}$ required for ohmic heating 
in the convection zone to be 30\% of the luminosity as estimated from 
no-feedback planet models, and compare with the results of our time-dependent 
calculations with feedback included. The curves show the 
$B_{\phi 0}$--$T_{\rm eq}$ predicted by the wind zone model for three 
different values of $B_r$. Black points show planets whose radii can be 
explained by our model ($R_{\rm pred}>R_{\rm obs}$ in Fig.~\ref{f:RPredict}), 
red points show planets that cannot be explained 
($R_{\rm pred}<R_{\rm obs}$ in Fig.~\ref{f:RPredict}). 
For clarity, we use the following abbreviations for planet names: 
W--WASP; H--HAT-P; K--Kepler; OG--OGLE-TR; C--CoRoT-P.
\label{f:hj}
}
\end{figure*}

To look in more detail at the effect of our assumed wind zone model on how 
successfully we are able to reproduce the observed radii, in Figure~\ref{f:hj} 
we show the results in the $B_{\phi 0}$--$T_{\rm eq}$ parameter space. For 
each observed planet, we first make a model with no ohmic heating, varying 
the internal entropy $S$ at the measured $M$ and $T_{\rm eq}$ until we match 
the measured radius $R_p$. Then we calculate the value of $B_{\phi 0}$ 
required in that model for the ohmic power in the convection zone $P_{\rm ohm}$ 
to be 30\% of the planet's luminosity. We colour-code the data points 
according to whether they are successfully explained by our time evolutions, 
ie.~whether they have $R_{\rm pred}$ larger or smaller than $R_{\rm obs}$ in 
Figure~\ref{f:RPredict}. These two groups of data points lie on either side of 
the $B_{\phi 0}$--$T_{\rm eq}$ relation from the wind zone model (solid curve). 
This shows that the approach of using a structural model with no ohmic heating (we refer 
to this as a ``no feedback'' model in \S \ref{Ty}) to estimate the critical magnetic field 
is a good approximation of our detailed time-evolution models including feedback. 

Comparing the red points in Figure~\ref{f:hj} with the solid curve gives a 
sense of how far short the ohmic heating model falls in explaining the most 
inflated planets. For example, HAT-P-32 is about a factor of $3$--$4$ above 
the curve, so that the heating rate ($\propto B^2$) needs to be increased by 
about an order of magnitude to explain the observed radius. It is interesting 
that most of the unexplained objects lie within a factor of 3 in terms of 
$B_{\phi 0}$ of the wind zone model. Figure~\ref{f:hj} helps to show what 
changes to the wind zone model would explain more of the observed objects. We 
have assumed the relation $T_{\rm iso}=1.33 T_{\rm eq}$ 
(from eq.~[\ref{eq:guillot}]); a larger factor between $T_{\rm iso}$ and 
$T_{\rm eq}$ would move the solid curve to the left, allowing ohmic heating to 
explain the radii of low $T_{\rm eq}$ planets such as WASP-06. The dashed and 
dotted curves show the effect of changing $B_r$. Increasing $B_r$ from 10 to 
100\,{\rm G} does increase $B_{\phi 0}$ at low temperatures, but reduces 
$B_{\phi 0}$ at high temperatures where magnetic drag is enhanced. A larger 
depth $\Delta\ln P$ would help to reduce the number of discrepant objects 
since both $B_{\rm adv}$ and $B_{\rm drag}$ increase with 
$\Delta\ln P$ (eqs.~[\ref{eq:Badv}] and [\ref{eq:Bdrag}]).


\section{Summary and Discussion} 
\label{discuss}

In this paper, we present models of ohmic heating in hot jupiters in which we 
attempt to decouple the interior and wind zone by replacing the wind zone by a 
boundary temperature $T_{\rm iso}$ and magnetic field $B_{\phi 0}$, both 
evaluated at a pressure $p=10$ bars. This approach allows us to survey the 
outcomes of ohmic heating, parametrized by $T_{\rm iso}$ and $B_{\phi 0}$ for 
planets with different mass $M$. This is similar in spirit to models of gas 
giant cooling, which often set an outer boundary pressure of 10\,bars and 
separately integrate a $T_{10}$--$T_{\rm eff}$ relation to use as an outer 
boundary condition.

The main conclusions of the paper are:

1. Figure~\ref{f:lpohm} is a key result, showing as a function of entropy 
how the ohmic power compares to the planet luminosity. Only planets with 
entropy below a critical value have enough ohmic heating to slow their 
contraction rate. Of particular note are the different mass dependences: 
at fixed $B_{\phi 0}$ and $T_{\rm iso}$, the cooling luminosity $L\propto M$ 
whereas the ohmic power decreases with mass 
(we find $P_{\rm ohm}\propto R^{2.4}/M$).

2. Ohmic heating has two effects on the thermal state of the planet. As well 
as providing direct heat input into the adiabatic convective interior
(as found by previous works, see \cite{B&S2010,Perna2010b,Wu2012}), the 
feedback of ohmic heating in the region between the wind zone and the 
convective boundary moves the convective zone boundary deeper 
(Fig.~\ref{f:pconv}), leading to a reduced cooling luminosity and reduced 
internal ohmic heating. Because the electrical conductivity changes 
dramatically with pressure through the planet, the total ohmic power inside 
the convection zone is very sensitive to its radial extent. To computing 
the planet age and radius at the late stage when ohmic heating halted the
cooling, it is crucial to accurately locate the convective-radiative boundary.

3. A larger $B_{\phi 0}$ is required for ohmic heating to be important in more 
massive planets or planets with larger $T_{\rm eq}$. This can be seen in 
Figures \ref{f:age} and \ref{f:Biso}, which show the age of a cooling gas 
giant when ohmic heating becomes important, and the magnetic field strength 
required for ohmic heating to be important at different values of 
$T_{\rm iso}$. For example, at a temperature $T_{\rm iso}=1750\ {\rm K}$, 
Figure~\ref{f:Biso} shows that $B_{\phi 0}\approx 30\ {\rm G}$ will halt the 
contraction of a $0.3\ M_J$ planet in 3 Gyr, whereas 
$B_{\phi 0}\approx 150\ {\rm G}$ is required for a $1 M_J$ planet at that temperature. 
At higher $T_{\rm iso}=2250\ {\rm K}$, the required values are 
$B_{\phi 0}\approx 100\ {\rm G}$ for a $0.3 M_J$ planet or  $B_{\phi 0}\approx 700\ {\rm G}$ 
for a $1 M_J$ planet.

4. With a specific model for the wind zone (\S \ref{sec:menou}), we can 
compare to observed systems as a function of their observed equilibrium 
temperatures $T_{\rm eq}$. The wind zone model specifies the induced field 
$B_{\phi 0}$ (or equivalently, the radial current that penetrates into the 
interior; see eq.~[\ref{eq:JR}]) as a function of $T_{\rm eq}$, and the 
relation between $T_{\rm eq}$ and the temperature at $10\ {\rm bars}$. Using 
the scaling analysis proposed by \cite{Menou2012} for the dynamics of the 
wind zone, together with the  atmospheric temperature profile from 
\citet{Guillot2010}, we find that it is difficult for ohmic heating to explain 
the large radii of hot jupiters with large masses and large $T_{\rm eq}$ 
(see Fig.~[\ref{f:RPredict}]). 

5. A more general approach is to calculate, for each observed planet, the 
$B_{\phi 0}$ that is required if ohmic heating is providing a significant 
fraction of the luminosity (and therefore able to significantly change the 
contraction rate of the planet). This is shown in Figure~\ref{f:hj} and shows 
how much the heating rate needs to be increased over the wind zone model in 
\S \ref{sec:menou} to explain particular objects. A modest increase in the 
wind zone thickness over that assumed here, or larger ratio of the temperature 
at depth $T_{\rm iso}$ compared to $T_{\rm eq}$, would improve the agreement 
with observed radii (see discussion in \S \ref{sec:obs}). Even so, several 
objects require a much more dramatic increase in heating rate 
(see Fig.~\ref{f:hj}).

The difficulty in explaining many of the observed radii that we have found 
differs from \cite{B&S2011} and \cite{Wu2012} who found that they could 
account for almost all of the observed hot jupiter radii with ohmic heating. 
The key difference is that we do not assume here that the heating efficiency 
(the fraction of the irradiation going into ohmic power, typically taken to be 
$\epsilon\sim 1$\%) to be fixed, but instead use the wind zone model to set 
the induced magnetic field in the wind zone and therefore the magnitude of the 
heating. 

It is important to emphasize that our conclusions about the efficacy of
ohmic heating depend on the particular prescription for the magnetic
field in the wind zone that we have used. In fact, many complexities
underlie the path from the irradiation to the properties of the induced
magnetic field. More realistic 3D wind zone models may give a different
picture than the simple 1D force balance scalings we have used here. For
example, in this paper we have assumed the wind zone extends to p = 10
bars. Figure 3 of \cite{Wu2012} nicely illustrates the importance
of the depth of the wind zone, showing that a shallower wind zone
requires a significantly larger overall efficiency to achieve the same
interior heating.  One situation in which this will break down is for
young planets with high entropies when the radiative/convective zone
boundary is at lower pressure. More work is needed on what happens when
the interior convection zone extends into the wind zone region.

Our results emphasize the key inputs that are necessary from atmospheric 
models: the thermal structure and dynamics of the wind zone including a large 
scale magnetic field, the values of induced magnetic field, or equivalently 
the magnetic Reynolds number $R_M$, that can be attained there, and the depth 
of the wind zone. More studies of the local conductivity profile 
and magnetic field properties in the high magnetic Reynolds number regime are 
needed. In particular, it is not clear whether the large values of induced 
field $B_{\phi0}>1000\ {\rm G}$ needed to explain the observed radii (Fig.~17) 
can be achieved in the wind zone. Furthermore, whether the implied large 
internal currents affect the planetary dynamo is also an open question.

Our results do compare favorably with previous calculations if we use 
equation (\ref{eq:nbphi}) to set a value of $B_{\phi 0}$ appropriate for the 
wind zone conditions assumed in those papers. For example, we are able to 
compute the $3\%$ heating profile at pressures $p>10\ {\rm bars}$ in Figure 4 
of \citet{B&S2010} by setting $B_{\phi0}=300\ {\rm G}$; we reproduce the 
heating profile of Tres-4b from \citet{Wu2012} with 
$B_{\phi0}\approx 1500\ {\rm G}$. However, a complication in comparing 
different models is that the heating dissipated in the wind zone is coupled 
with the heating dissipated in the planet interior. \citet{Wu2012} in 
particular discuss the expected ratio of heating deposited in different layers. 
But this ratio is generally model dependent and varies through the planet 
lifetime. A direct result of this kind of coupling is that models with same 
heating efficiency but different wind zone model are not physically 
comparable. For example, for a given set of planet properties, the radius 
predicted by \citet{B&S2010} is larger than in \citet{Wu2012} for the same 
choice of heating efficiency $\epsilon$, because the heating ratio between the 
wind zone and the interior is much smaller in \citet{B&S2010}, creating a much 
stronger internal heat source. Similarly, although \citet{Menou2012} estimated 
the total ohmic heating efficiency to be $>1\%$ over a certain range of 
equilibrium temperatures (with the weather layer between $60$ mbars and $150$ 
mbars), the internal heating has a much lower efficiency, 
consistent with our findings in \S \ref{observation}.

Another uncertainty is in the microphysics aspects of the electrical 
conductivity. For example, as we noted in \S 3.1, 
\cite{B&S2010} make a different choice for the electron-neutral cross-section 
and thermal averaging that results in a factor of 9 difference in electrical 
conductivity than we adopt here. The estimates in \S \ref{AR} show that the 
amount of ohmic power is sensitive to changes in the electrical conductivity 
(or the ionization fraction) in two ways. At low densities in the wind zone, 
the conductivity determines the size of the current (eq.~[\ref{eq:RM}]); 
in the interior, the ohmic power is $\propto 1/\sigma$ 
(eq.~[\ref{eq:pohmtotal}]). For a fixed efficiency $\epsilon$, a different 
normalization for $\sigma$ does not change the evolution of the planet, since 
the normalization of the heating profile is determined by the choice of 
$\epsilon$, and $\sigma(r)$ determines only its shape. The normalization of 
$\sigma$ is important, however, when going beyond the constant efficiency 
assumption, making it crucial to understand the processes that set the 
ionization level in hot jupiter atmospheres.

\begin{figure}
\includegraphics[angle=0,width=\linewidth]{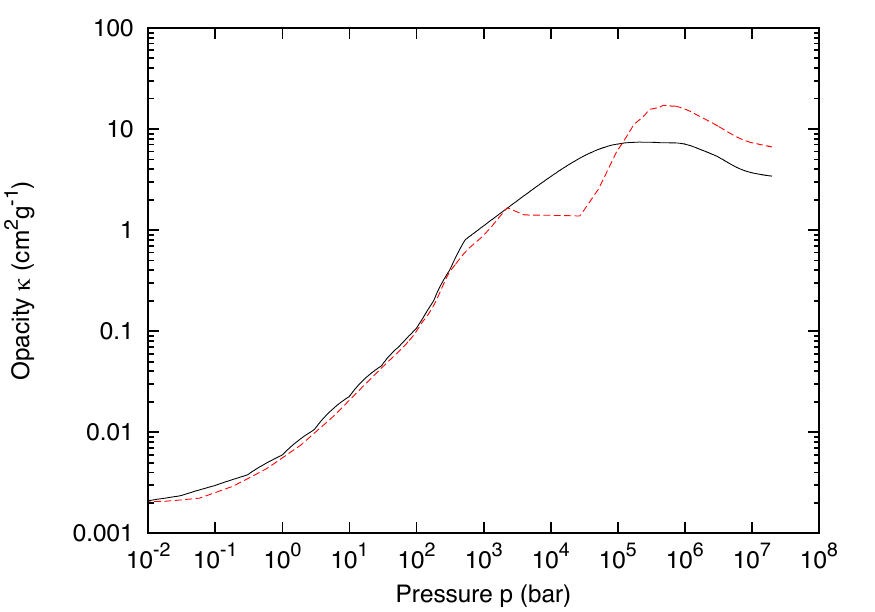}
\caption{
The opacity profile in a planet with parameters as in Table 1 tablenote a (no 
ohmic heating, radiative model), showing the opacity as calculated by combining 
the \citet{Freedman08} and \citet{Potekhin2010} tables (solid curve) 
or from the MESA code \citep{Paxton11} (red dashed curve). 
\label{f:opacity}
}
\end{figure}

\begin{figure}
\plotone{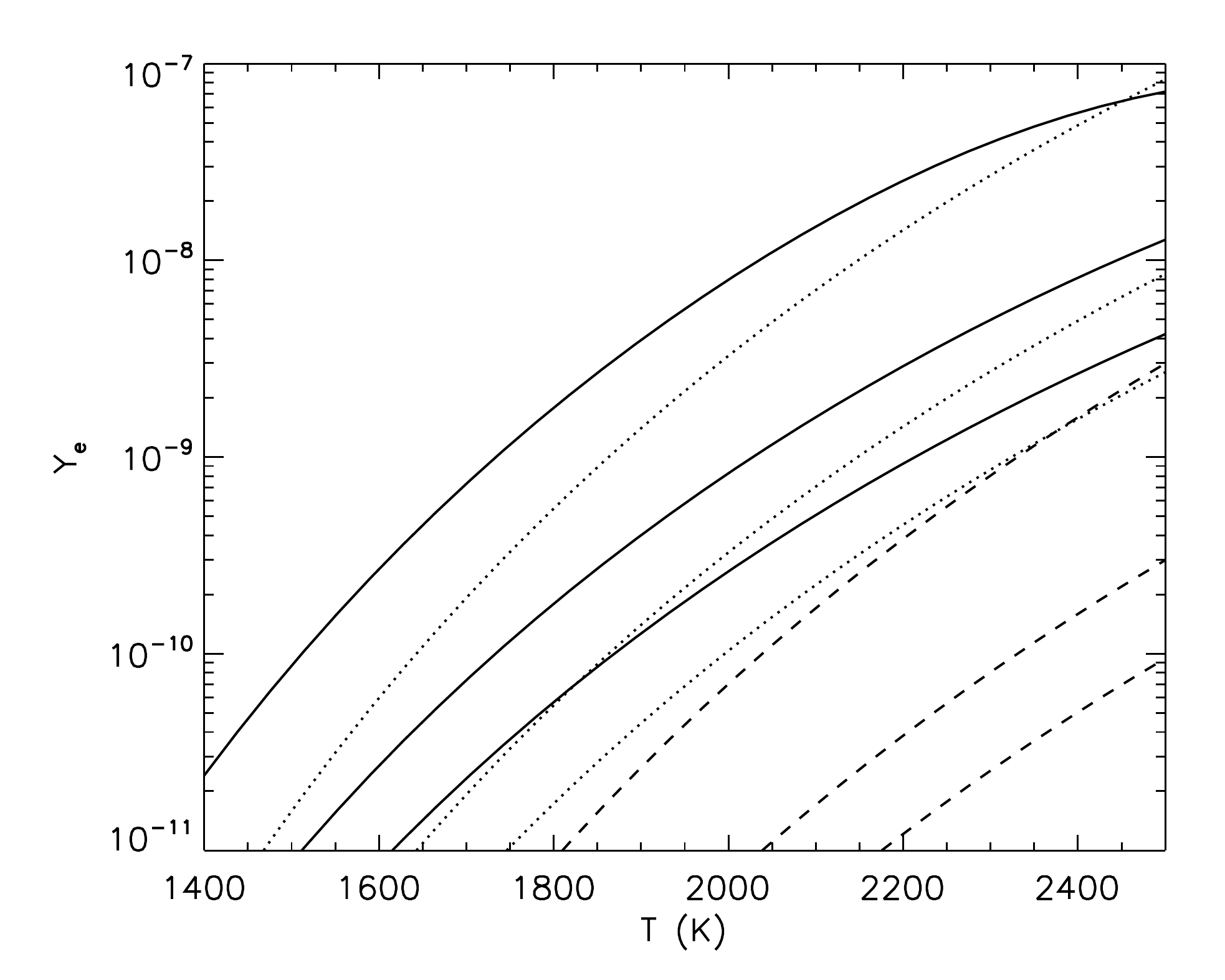}
\caption{
The contribution to the electron fraction $Y_e$ from different alkali metals 
as a function of temperature and pressure. Solid curves are for potassium, 
dotted for sodium, and dashed for aluminum. In each case, we show (top to 
bottom) pressures of 1, 100 and 1000 bars.
\label{f:Yep}
}
\end{figure}

\acknowledgements{This work began as a project at the 2011 International 
Summer Institute in Modeling in Astrophysics (ISIMA), held at the Kavli 
Institute of Astronomy and Astrophysics, Beijing, China. We thank ISIMA
for support and KIAA for hospitality during the program.  
We would like to thank E.~Chiang, D.N.C.~Lin, A.~P.~Showman, Y.~Wu and 
Y.~Lithwick for useful discussions during 2011 ISIMA. We are also grateful for 
the helpful suggestions during private communication from T.~Guillot and 
R.~Laine, and to G.-D.~Marleau for discussions about gas giant models and 
a thorough reading of the paper. AC is supported by an NSERC Discovery Grant.}

\appendix
\section{Microphysics of the planet interior}
\label{sec:microphysics}

We discuss the microphysics input in our gas giant models here. We adopt the 
equation of state from \citet{Saumon 1995} with helium fraction $Y=0.25$. In 
order to maximize the planet radius, we do not include a solid core or 
elements heavier than helium.

The radiative opacity is taken from \cite{Freedman08} and in the core we
include thermal conduction by electrons from \cite{Potekhin2010}. The
transition from radiative to conducive opacity occurs at a pressure
which is greater than the maximum pressure of 300 bars covered by the
\cite{Freedman08} tables. In the intermediate regime, we assume the
scaling $\kappa\propto{p}^{0.5}$. The opacity profile for our standard
model is shown in Figure~\ref{f:opacity}, over-plotted with opacity
taken from MESA using the same planet structure \citep{Paxton11}. 
 
\begin{figure}
\epsscale{1.1}
\plotone{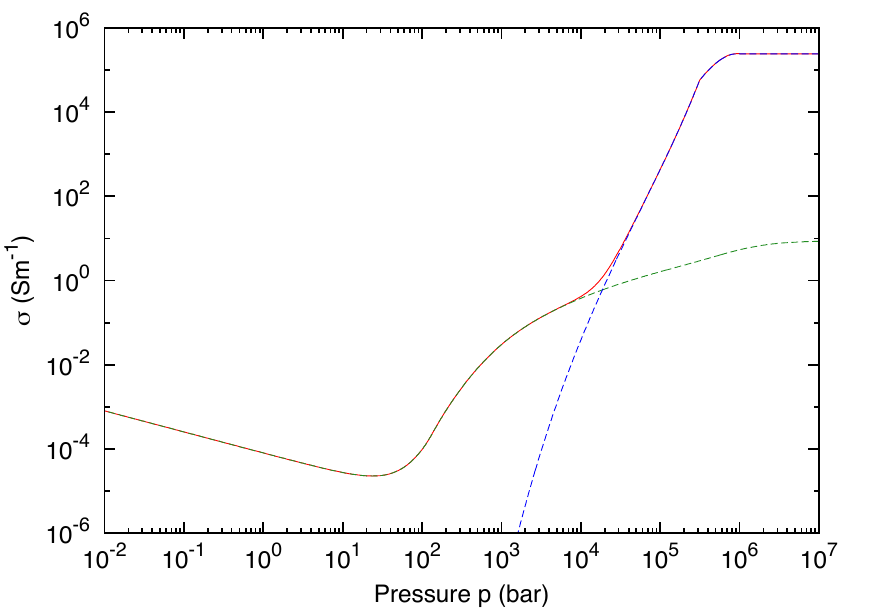}
\plotone{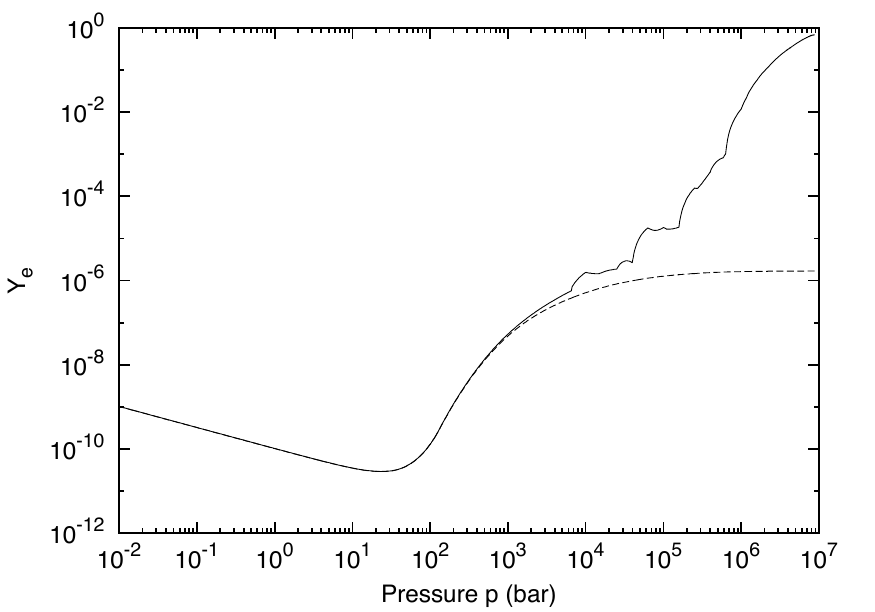}
\caption{
Top panel: the electrical conductivity profile of a planet with parameters as 
in Table 1 tablenote a (no ohmic heating, radiative model), showing the 
contributions from alkali metals (dashed green curve) and  hydrogen (dashed 
blue curve).  Bottom panel: the electron fraction $Y_e$ as a function of 
pressure, with the contribution from alkali metal ionization shown as a dashed 
curve.
\label{f:sigma}
}
\end{figure}

The electrical conductivity has contributions from alkali metal ionization in 
the outer layers, and hydrogen in the interior. In the upper atmosphere of hot 
jupiters, the conductivity is set by the ionization of alkali metals. For 
potassium, which has the lowest ionization potential
\footnote{The first ionization potentials of K, Na, Al, Mg and Fe are 4.34, 5.14, 5.99, 7.65 and 7.90~eV respectively \citep{David 2003}.}. 
The potassium only Saha equation \citep{Balbus2000,Perna2010a} 
gives the ionization fraction $x_k=n_e/n$ as 
\begin{equation}
x_k=\left[\frac{f_k}{n}\left(\frac{m_ek_BT}{2\pi\hbar^2}\right)^{3/2}e^{-4.35 {\rm eV}/k_BT}\right]^{\frac{1}{2}}
\label{eq:saha}
\end{equation}
\begin{displaymath}
=1.03\times10^{-3}\ T_3^{5/4}e^{-25.19/T_3}
\left(\frac{f_K}{10^{-7}}\right)^{1/2}
\left(\frac{p}{1 {\rm bar}}\right)^{-1/2},
\end{displaymath} 
where $f_K$ is the number fraction of potassium.
Although potassium dominates, we also include the contribution of 
Na, Mg, and Fe in the ionization balance to sum up the total ionization fraction. 
The ionization fraction of each alkali metal is computed separately 
by assuming a balance independent on the presents of other elements . 
We do not include the contribution of Al in the
calculation, which is likely condensed out \citep{Lodders1999}.
But our results are not very sensitive to elements beyond potassium. 
This is illustrated in Figure~\ref{f:Yep} which shows the contributions to the 
ionization level from K, Na and Al at different pressures. Once the ionization fraction is 
determined, the conductivity is $\sigma=n_e e^2/m_e\nu$ where the collision 
frequency of electron-neutral collisions is 
$\nu_{\rm en}=n_n\langle\sigma{v}\rangle_e$ given by \citet{Draine83} as
\begin{equation}\label{eq:sigv}
\langle\sigma{v}\rangle_e=10^{-15}\left(\frac{128k_BT}{9\pi{m_e}}\right)^{1/2}\
{\rm cm}^3\ {\rm s}^{-1}.
\end{equation} 
The conductivity is then 
\begin{equation}
\sigma=8.8\times 10^{-2}\ {\rm S\ m^{-1}}\ \left({x\over 10^{-7}}\right)\left({T\over 1500\ {\rm K}}\right)^{-1/2}.
\end{equation}

In the deeper part of the planet, the hydrogen is ionized by high pressure and 
the conductivity is dominated by electron-proton collisions. In the 
fully-degenerate limit, 
$\nu_{\rm epd}=4e^4m_e\Lambda/3\pi\hbar^2=1.8\times\,10^{16}\,{\rm s}^{-1}$. 
In the non-degenerate limit, 
$\nu_{\rm epnd}=6.4\times 10^{23}\ s^{-1}\rho{Y_e}T^{-3/2}$, in which $Y_e$ is 
the electron fraction. We interpolate between the two limits to give an 
estimation of the total contribution: 
$\nu_{\rm ep}^{-2}=\nu_{\rm epd}^{-2}+\nu_{\rm epd}^{-2}$. We also include the 
conductivity at intermediate densities as given by \cite{Liu thesis}. Before 
the hydrogen molecule is fully ionized, the band-gap of hydrogen will diminish 
with increasing pressure, to a level where there is a significant contribution 
to the conductivity. \cite{Liu thesis} give this as
\begin{equation}
\sigma_s= \sigma_0 \exp\left(\frac{-E_g(\rho)}{k_BT}\right)
\end{equation}
where between 0.2 Mbars and 1.8 Mbars, $E_g=20.3-64.7\rho$, where $E_g$ is in 
eV, and $\rho$ is in ${\rm mol}\,{\rm cm}^{-3}$, and 
$\sigma_0=3.4\times10^{20}\ \mathrm{s}^{-1}\ \exp(-44\rho)$. The overall 
conductivity is constructed as 
$\sigma=\sigma_s+n_e\,e^2/m_e\nu=\sigma_s+1.52\times10^{32}\rho{Y_e}/\nu$. 
The collisional frequency $\nu$ is the sum of electron-neutral and 
electron-proton collisions. A typical conductivity profile and the contribution 
of different components are shown in Figure~\ref{f:sigma}.

\clearpage

\end{document}